# Impact IRR: Leveraging Modern Portfolio Theory to Define Impact Investments

Daniel Soliman

Daniel Soliman
is the principal at Pareto Impact Ventures in Chapel Hill, NC.
dan@piv.group

## KEY FINDINGS

- The author proposes impact IRR as a framework for utilizing components of modern portfolio theory, financial tools, and existing data to evaluate and monitor impact investments.
- The author demonstrates the technical implementation of impact IRR to evaluate a potential investment.
- The author demonstrates how to utilize impact IRR through three illustrative use cases from leading impact investment funds.

## ABSTRACT

The impact investment market has an estimated value of almost $1.6 trillion. Significant progress has been made in determining the financial returns of impact investing. Investors are still, however, in the early stages of determining impact return. In this study, the author proposes the use of impact internal rate of return (impact IRR) to evaluate and monitor impact investments. This approach, which utilizes components of modern portfolio theory, adapted financial tools, and existing datasets, is demonstrated herein through initial use cases and examples showing how it can be employed to optimize impact.

## OVERVIEW

### The Problem and Opportunity

"Doing well by doing good," a principle attributed to Benjamin Franklin, aptly captures the essence of impact investing. Investors have become increasingly aware of the potential to address global challenges through strategic investments. As more investors seek to balance financial gains with societal benefits, the impact investing market continues to expand, driving meaningful change across diverse sectors.

The Global Impact Investing Network (GIIN) estimates that the impact investment market is valued at almost $1.6 trillion (Hand, Ulanow, et al. 2024). The industry has made great strides in determining the financial return of impact investments. It is, however, in the early stages of answering the question, *What is an impact return?*[1]

[1] The lack of a standardized and measurable methodology for evaluating impact returns has been cited as one of the industry's most significant weaknesses (The Brainy Insights 2024).

Leaders such as Impact Frontiers (IF) are working to establish consensus regarding norms and best practices. First-mover GIIN established IRIS+ as a means of measuring and managing impact. In the absence of established practices, however, the current impact investing monitoring and evaluation environment is a veritable Wild West of excitement, innovation, and uncertainty as the sector works to address four key questions:

1. What is impact?
2. How do we measure impact?
3. What is the impact return of an impact investment?
4. How can impact investments fund this information effectively to their stakeholders?

There are two currently underutilized areas to standardize impact evaluation:

1. Existing financial tools, methodologies, and concepts.
2. Data available to the impact investors.

The objective of this study is to provide investors with tools to determine the impact return of an investment. This objective will be accomplished by introducing and applying the concept of impact internal rate of return (impact IRR) to formalize the impact investment evaluation process and by updating an existing financial tool—the internal rate of return/net present value (IRR/NPV) formula—so that investors can effectively evaluate opportunities to invest in social and environmental solutions in a way that can be incorporated into existing financial due diligence and evaluation processes.

Impact IRR standardizes the evaluation of data by focusing on the monetized components of an outcome or impact. It is intended to be just one facet of the evaluation of an investment, analogous to how IRR is just one aspect of conventional assessments.

This study will focus on:

1. Understanding the impact and financial risks and returns of an investment.
2. Identifying an investment with a projected positive intended outcome or impact.
3. Monitoring financial and impact returns throughout the investment term.

This study will explain impact IRR and make the case for using it to assess investments. Then, it will define initial use cases and present illustrative use cases that incorporate impact IRR into realistic scenarios for impact investment funds.

## Impact Investing Definitions

GIIN defines impact investments as investments made with the intention to generate positive, measurable social and environmental impact alongside a financial return. Impact investments have four core characteristics: intentionality, use of evidence and impact data in investment design, managing impact performance, and contributing to the growth of the industry (Global Impact Investing Network 2019).

Impact investments are only one category within the social/environmental funding continuum (Zhou 2022). See Exhibit 1.[2]

Generally, there are two types of impact investment capital:

1. *Market-rate impact capital (MIC)*: investments that seek to achieve financial returns equal to those of comparable nonimpact investments while also intending to generate a positive and measurable social or environmental outcome or impact.
2. *Below-market-rate impact capital (BIC)*: investments that provide the necessary incentives for market-rate investors to participate in an investment structure intended to yield a measurable social or environmental outcome or impact.

## EXHIBIT 1
### Continuum of Capital

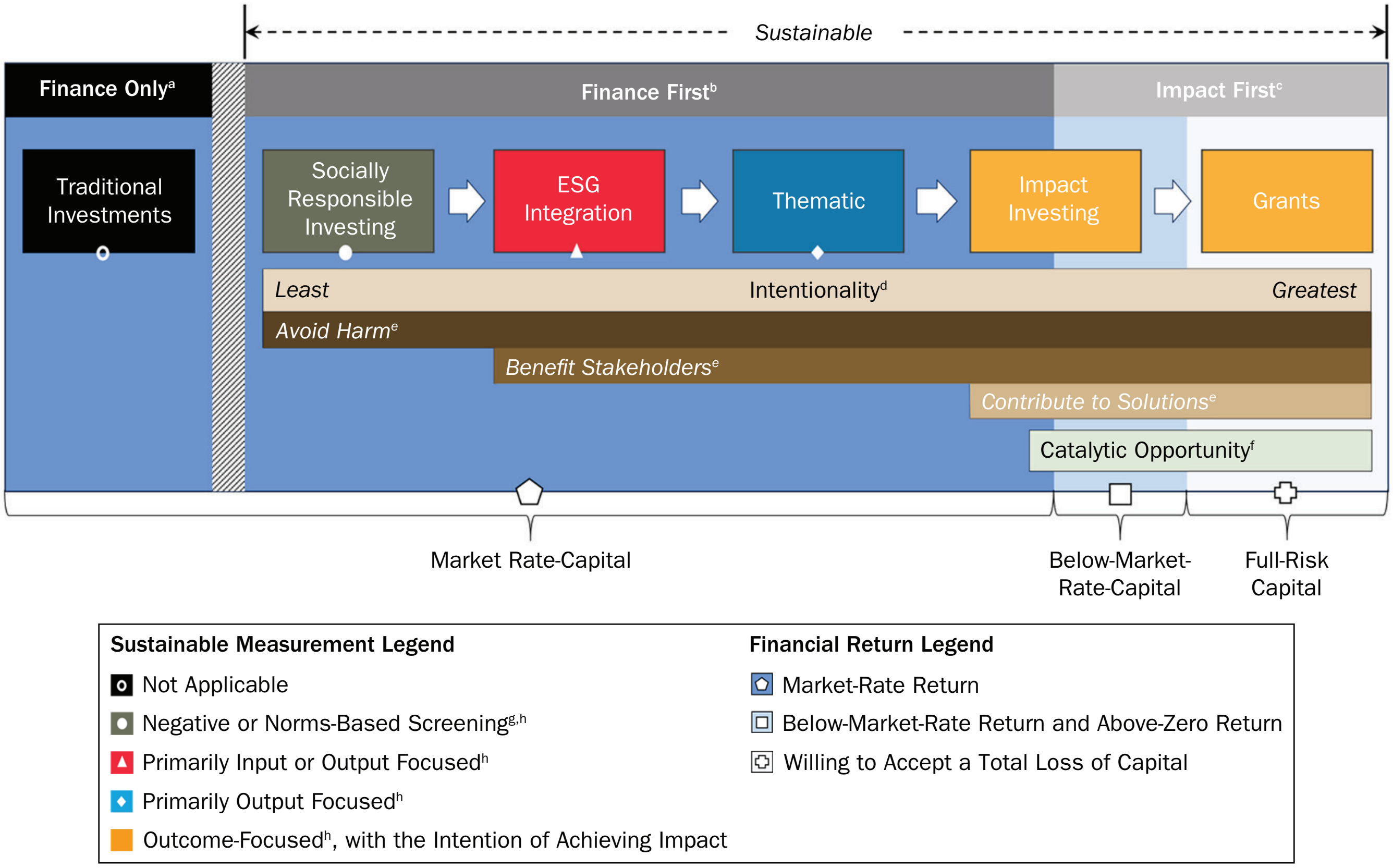

**NOTES:** [a]Exclusively focused on financial returns. [b]Prioritize financial over impact returns. [c]Prioritize impact over financial returns. [d]Intention to contribute to measurable positive social or environmental benefit (Global Impact Investing Network, n.d.). [e]Impact investing requires all three (Impact Frontiers, n.d.). [f]Investments designed to generate or enhance impact by attracting market-rate investors who would not otherwise invest. For a complete description, see page 17 ("Element 7: Catalytic Opportunity"). [g]Wagstaff, Revesz, and Stewart (2024). [h]Impact intentionality measurement metrics derived from Tideline's Framework for Impact Labeling. (BlueMark and Morgan, Lewis & Bockius LLP 2021).

[2]The sector is witnessing a surge of new designations that could change the sustainable finance continuum. For example, the United Kingdom's Financial Conduct Authority's sustainability disclosure requirements clarify the degrees of indirect and direct impact used for the classification of portfolios (J. P. Morgan Asset Management 2024). It will be essential to monitor these new designations to assess their potential effects on the current continuum and impact investing.

BIC is only necessary for potential investments when financial concessions (1) induce market-rate investors to commit capital by creating a viable investment structure, and (2) establish criteria that ensure the investment generates or enhances the specified outcomes or impact goals set by the BIC investor(s).

BIC may include guarantees for debt at below-market rates, and equity with asymmetrical returns (Convergence 2018). Other names for BIC are patient capital (Acumen, n.d.), concessionary capital, and catalytic capital[3] (MacArthur Foundation, n.d.). Program-related investments (PRI) are a type of BIC (Internal Revenue Service 2023).

## LITERATURE REVIEW

This article contributes to the impact investing literature in three areas: impact measurement and data, evaluation of impact investment returns, and definition of investor preferences while optimizing impact.[4]

First, this article contributes to the literature by introducing a measurement framework based on modern portfolio theory that integrates a theory of change and repurposes existing datasets to measure impact. A key theme in the literature is the current limitations in measuring impact and data. In their systematic review of the impact investing literature, Schlütter et al. (2024) note that impact investing lacks standardized measurement approaches, has unclear metrics, and is overly complex, leading to a reliance on storytelling and qualitative evaluations. Similarly, Feor, Clarke, and Dougherty (2023), in their systematic literature review of social impact measurement, found a lack of standardization and cumbersome implementation among existing models in the assessment of social impact in impact investing. All but two of the items[5] presented in this review indicate that impact measurement and data are limiting factors or areas for improvement.[6] Furthermore, in GIIN's *State of the Market 2024* report, 92% of respondents indicated that impact measurement and management were a challenge (Hand, Sunderji, et al. 2024).

Second, this article contributes to the literature by introducing a tool for evaluating financial and impact returns within the previously mentioned measurement framework. The literature evaluating the performance of impact investment funds is limited; it focuses primarily on comparisons between market-rate impact funds and nonimpact funds.[7] Jeffers, Lyu, and Posenau (2024) developed a new parameter to evaluate the risk characteristics of private market asset classes, reporting that matched nonimpact funds demonstrated higher absolute performance than impact funds. After adjusting for market risk exposure, however, the performances across strategies were more aligned with each other. Barber, Morse, and Yasuda (2021) found that impact funds earned an

[3] There are some instances in which catalytic capital is priced at market rates in acknowledgment of the higher risk associated with being a first mover in the space. For a complete description of catalytic capital, see page 17 ("Element 7: Catalytic Opportunity").

[4] Distinguishing between impact investing research can be challenging because many researchers use the terms *ESG*, *sustainable investing*, and *impact investing* interchangeably. This review will focus on impact investing as previously defined by GIIN and highlight relevant literature from other areas of sustainable investing.

[5] Both Broccardo, Hart, and Zingales (2021) and Gupta, Kopytov, and Starmans (2024) focus on exclusion/disinvestment strategies.

[6] Chowdhry, Davies, and Waters 2018; Cole et al. 2020; Berk and van Binsbergen 2021; Oehmke and Opp 2024; Barber, Morse, and Yasuda 2021; Edmans, Levit, and Schneemeier 2023; Gupta, Kopytov, and Starmans 2024; Landier and Lovo 2020; Lo and Zhang 2023; Green and Roth 2024; Jeffers, Lyu, and Posenau 2024.

[7] Jeffers, Lyu, and Posenau (2024) give data limitations as one reason for the lack of research; the research cited focuses on impact investment private equity funds, the second largest impact investment allocation; private debt is the largest (Hand, Sunderji, et al. 2024). At least one article stated that private debt had not been researched because they could not find data sources (Barber, Morse, and Yasuda 2021).

IRR 4.7% lower than matched nonimpact funds. Conversely, Cole et al. (2020) cited a prominent impact investor who had achieved an IRR 15% above the market.

Third, this article contributes to the literature by exploring a method to determine impact investors' preferences for both financial and impact returns. The literature on investor preferences and maximizing financial and impact returns focuses on various sustainable investment types.[8] Barber, Morse, and Yasuda (2021) developed a willingness-to-pay model and observed that impact investors accepted IRRs 2.5 to 3.7 percentage points lower for impact investing funds compared to nonimpact funds. Lo and Zhang (2023) developed an order statistics framework to assess the financial impact of socially and environmentally focused investments. They offered a method for ranking such investments and presented several case studies, including one impact investment. Oehmke and Opp (2024) introduced a social profitability index to assess the effectiveness of green ESG funds at generating impact. Green and Roth (2024) presented a framework to explore how investors influence social outcomes through asset allocation decisions, demonstrating that equilibrium asset allocation varied significantly depending on how the investors defined social value.[9]

## RATIONALE FOR IMPACT IRR/IMPACT NPV AND INITIAL USE CASES

### Framing Impact Investing

Risk and reward are two foundations of modern portfolio theory. The risk–reward relationship is defined through the efficient frontier: the set of optimal portfolios that offer the highest expected return for a defined level of risk or the lowest risk for a given level of expected return (Goetzmann, n.d.).

Building on the efficient frontier concept, IF developed the efficient impact frontier (EIF), a set of portfolios that offer optimal combinations of impact *and* financial performance (i.e., portfolios optimized for risk-adjusted financial returns and impact returns) (Impact Frontiers, n.d.).

To develop the range of impact investments using the efficient impact frontier, I constructed a chart depicting the risk-adjusted impact returns on the *x*-axis and risk-adjusted financial returns on the *y*-axis (see Exhibit 2). Within this range of investments, two thresholds emerge:

1. *Impact threshold*: The minimum impact required for an investment to be classified as an impact investment. Investments that do not meet this requirement are either traditional investments or non-investable investments (do not satisfy either the minimum required traditional financial or impact returns).
2. *Financial threshold*: There are two investment types within the impact investment range:
   - *Market-rate impact investments*: investments that meet the minimum required market return.
   - *Below-market-rate impact investments*: investments with lower projected returns than the minimum required for market-rate capital.[10]

A third type of impact investment, blended capital, emerges when BIC and MIC are combined (see Exhibit 3). Blended finance combines capital with different levels

[8] Additional research examining investor preference for unsustainable investments to produce impact includes Chowdhry, Davies, and Waters (2018), Berk and van Binsbergen (2021), Broccardo, Hart, and Zingales (2021), Edmans, Levit, and Schneemeier (2023), Gupta, Kopytov, and Starmans (2024), and Landier and Lovo (2020).

[9] Their model examines investors who utilize both MIC and BIC investments.

[10] Capital with a negative projected financial return is considered a grant.

## EXHIBIT 2

### Range of Impact Investments

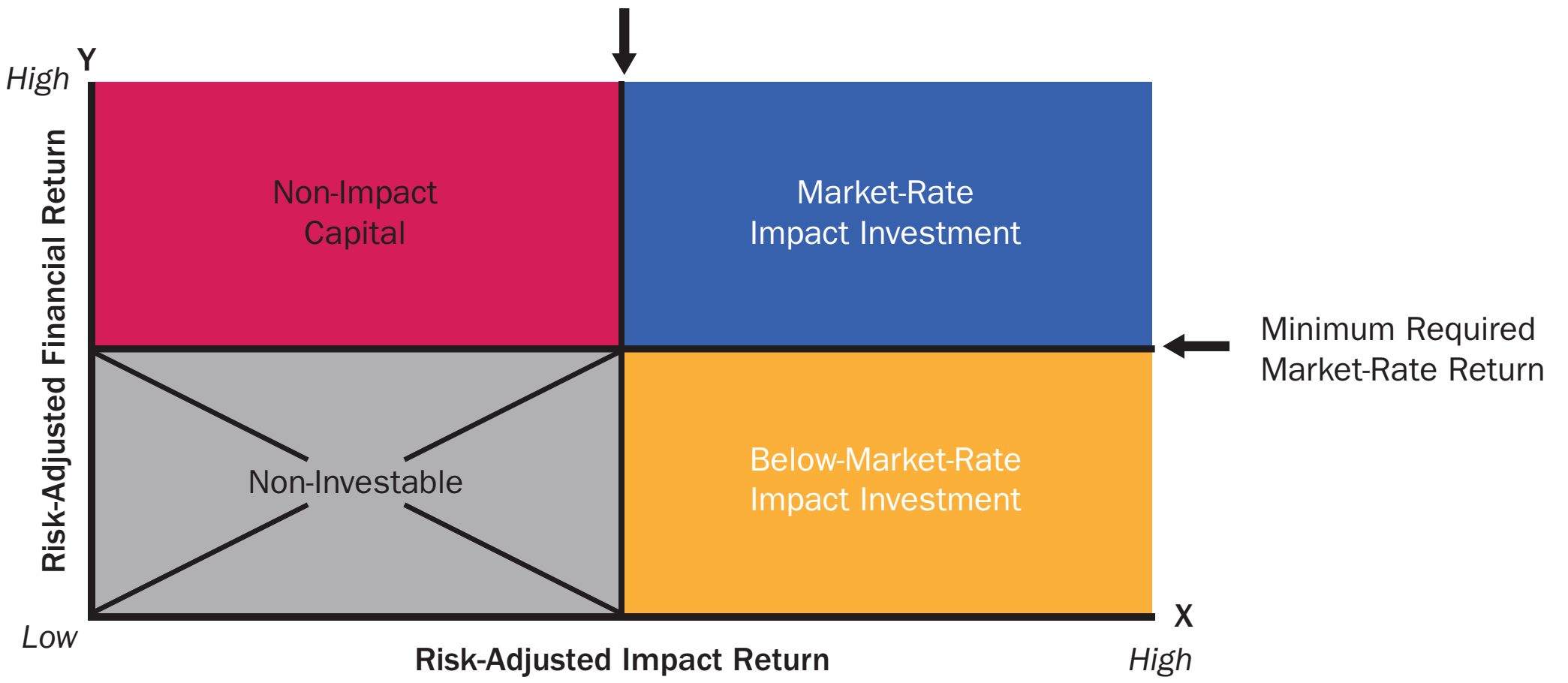

## EXHIBIT 3

### Complete Range of Impact Investments

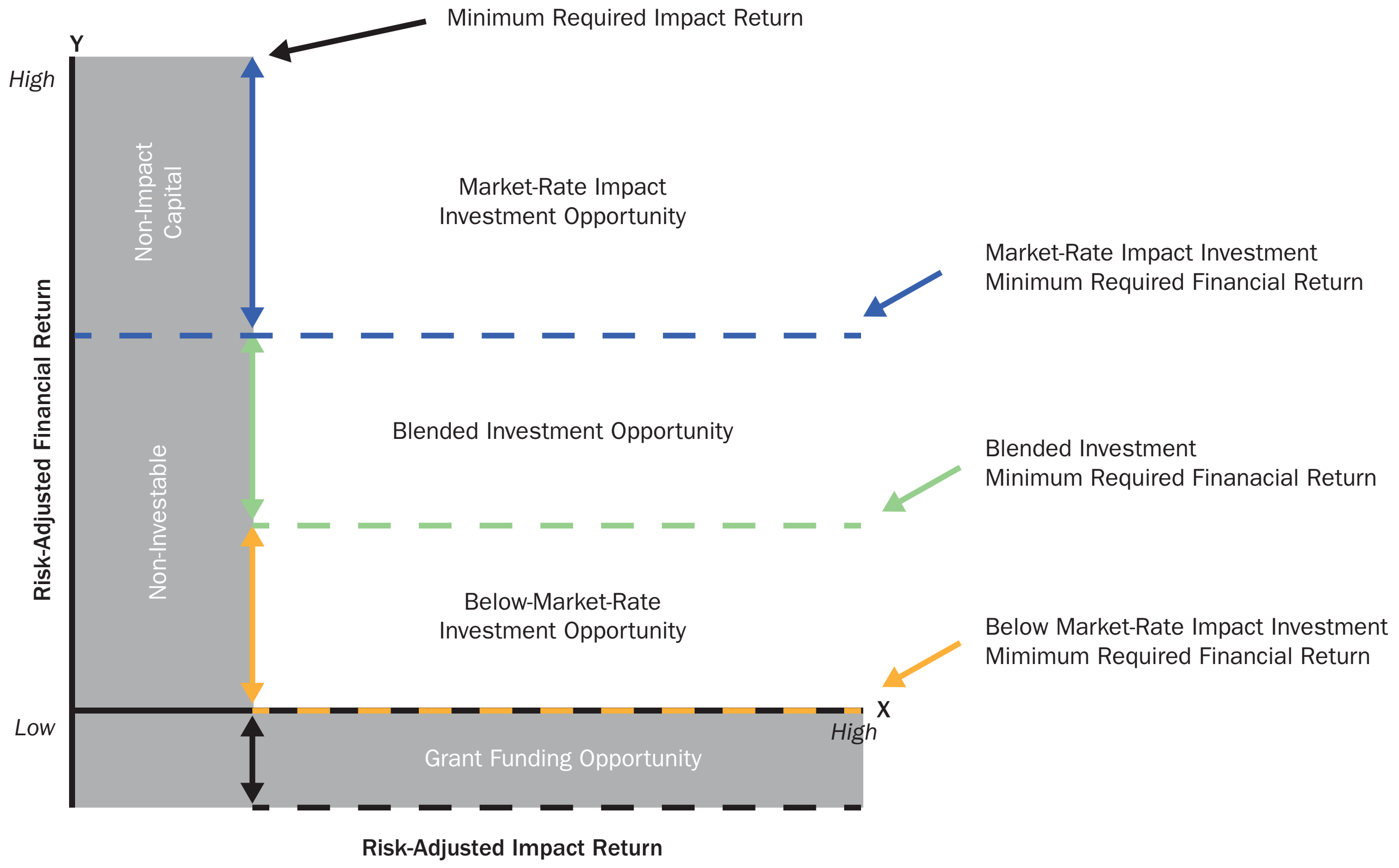

of risk to attract risk-adjusted market rate–seeking financing to impact investments (Global Impact Investing Network, n.d.).[11]

If impact investors shared data for determining the risk–reward profiles of comparable investments, they could better allocate funds to the appropriate capital type: MIC, BIC, and, if applicable, grant funding[12] (see Exhibit 4).

Further, the sharing of such data would optimize the sector to answer three essential questions:

1. *Investor level*: What are the projected financial and impact returns for an impact investment?
2. *Sector level*: What are the optimal valuation bands for MIC and BIC across each asset class and impact theme? When is BIC most beneficial for creating or scaling impact?

## EXHIBIT 4
### Systemic-Level Impact Investment Opportunities Range

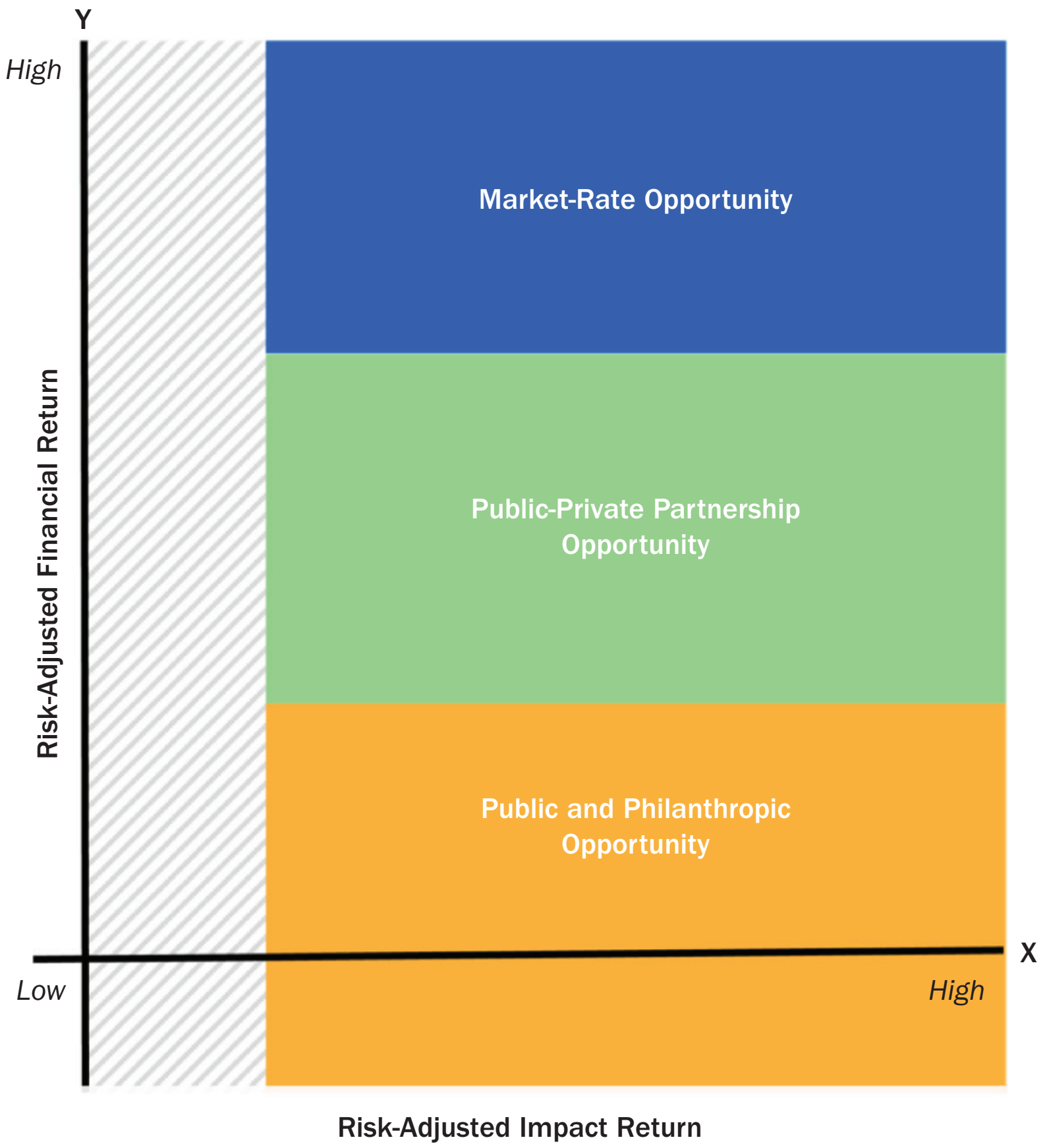

[11] There are two blended finance strategies: (1) MIC and BIC are mutually exclusive within the investment structure (e.g., a real estate project in which all sources are separate within the capitalization exhibit), and (2) MIC and BIC are combined to make investments (e.g., a private equity fund that has both MIC and BIC investors). In both cases, the BIC is priced to induce the MIC investor to participate, and the weighted average cost of capital is reduced to make the investments viable.

[12] Grant funding in this context is defined as capital in which an investor will focus solely on impact return and is willing to have a 100% negative financial return. Grant funding is not an impact investment opportunity.

3. *Systems level*: Capital is allocated into three investment opportunities:[13]
   a. *Market-rate investments*: investments made to achieve both commercial and intended impact returns
   b. *Public–private investments*: investments that pool private-sector capital that seeks a market-rate return with public-sector capital (from central and local governments or publicly owned corporations) that is priced below market rates to achieve an acceptable blended financial return and intended impact returns
   c. *Public and philanthropic investments*: investments made by the public sector or private (for-profit and nonprofit) entities that are priced below market rate and prioritize intended impact returns

## Acceptable Impact Return: A Missing Piece of the Puzzle

To optimize impact, investors need to answer four questions:

1. How is impact defined?
2. How is impact measured?
3. When is impact measured?
4. What is an acceptable level of return for impact?

Impact return is a developing field with varying models for projecting returns. Examining the concepts of the efficient frontier and the EIF provides insight into developing a measure of return on impact.

Exhibit 5, Panels A and B, depict an optimal investment/portfolio through two graphs, each with return on the *y*-axis and risk on the *x*-axis. Panel A shows the efficient

### EXHIBIT 5
**Efficient Frontiers**

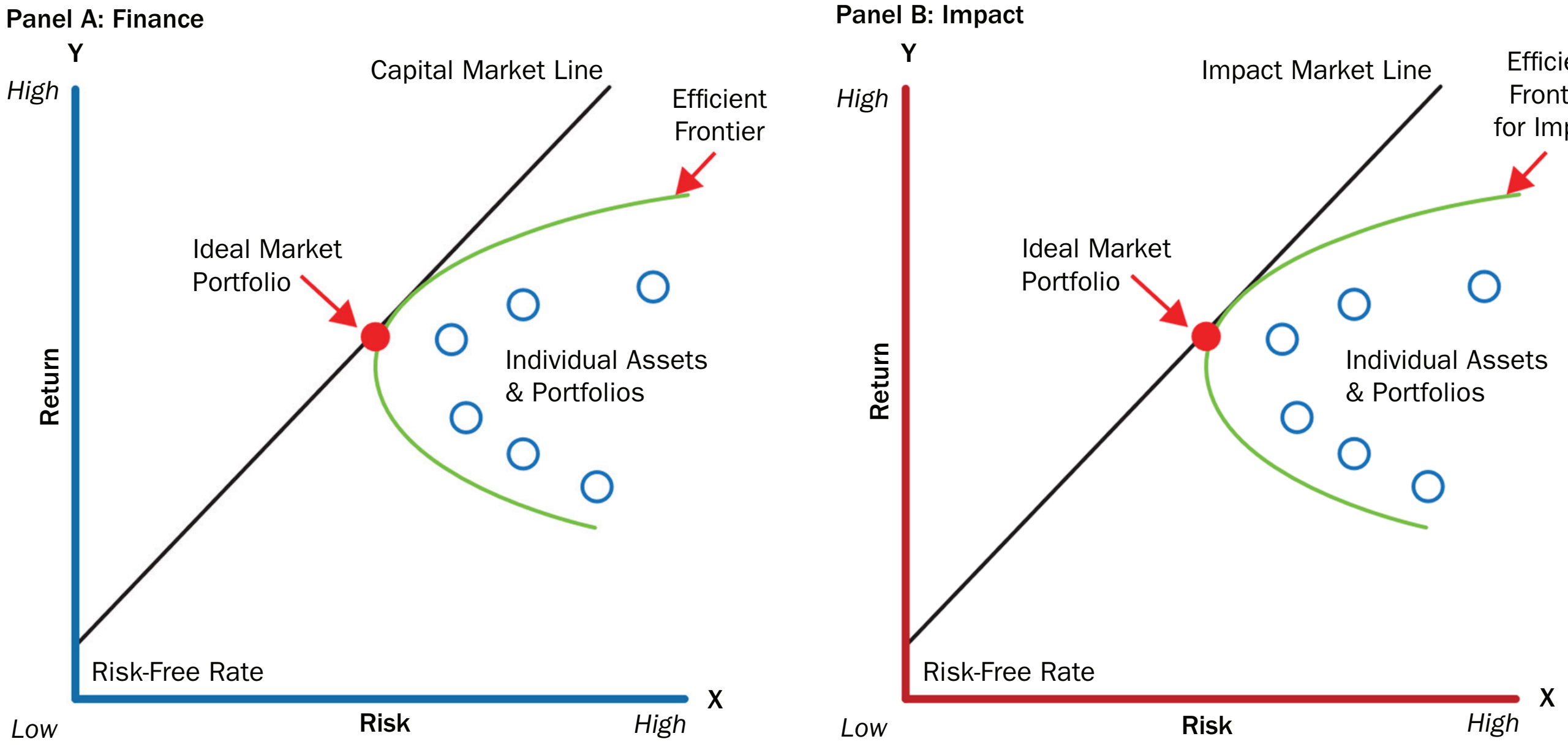

[13] Once comprehensive datasets encompassing a critical mass of investments across impact asset classes and themes—capturing both financial and impact return characteristics—have been established, subsequent research should examine how to optimize financial return and impact performance. Thus, suitable impact valuation bands spanning the investor spectrum could be determined.

frontier, which defines the area within which the *financial* return of an investment will be maximized for a given level of risk. Panel B shows the same thing, but for *impact*. Impact investors currently only have a standardized way of evaluating financial return using IRR/NPV.

The financial and impact risks and returns are represented graphically by constructing an opportunity field that employs a *z*-axis connecting two parallel planes representing the two graphs (see Exhibit 6), thereby creating the EIF. The EIF illustrates the opportunity set for all impact investments.[14] Representing EIF in this three-dimensional space makes it possible to determine an acceptable return for an impact investment.[15]

## EXHIBIT 6

### Efficient Impact Frontier

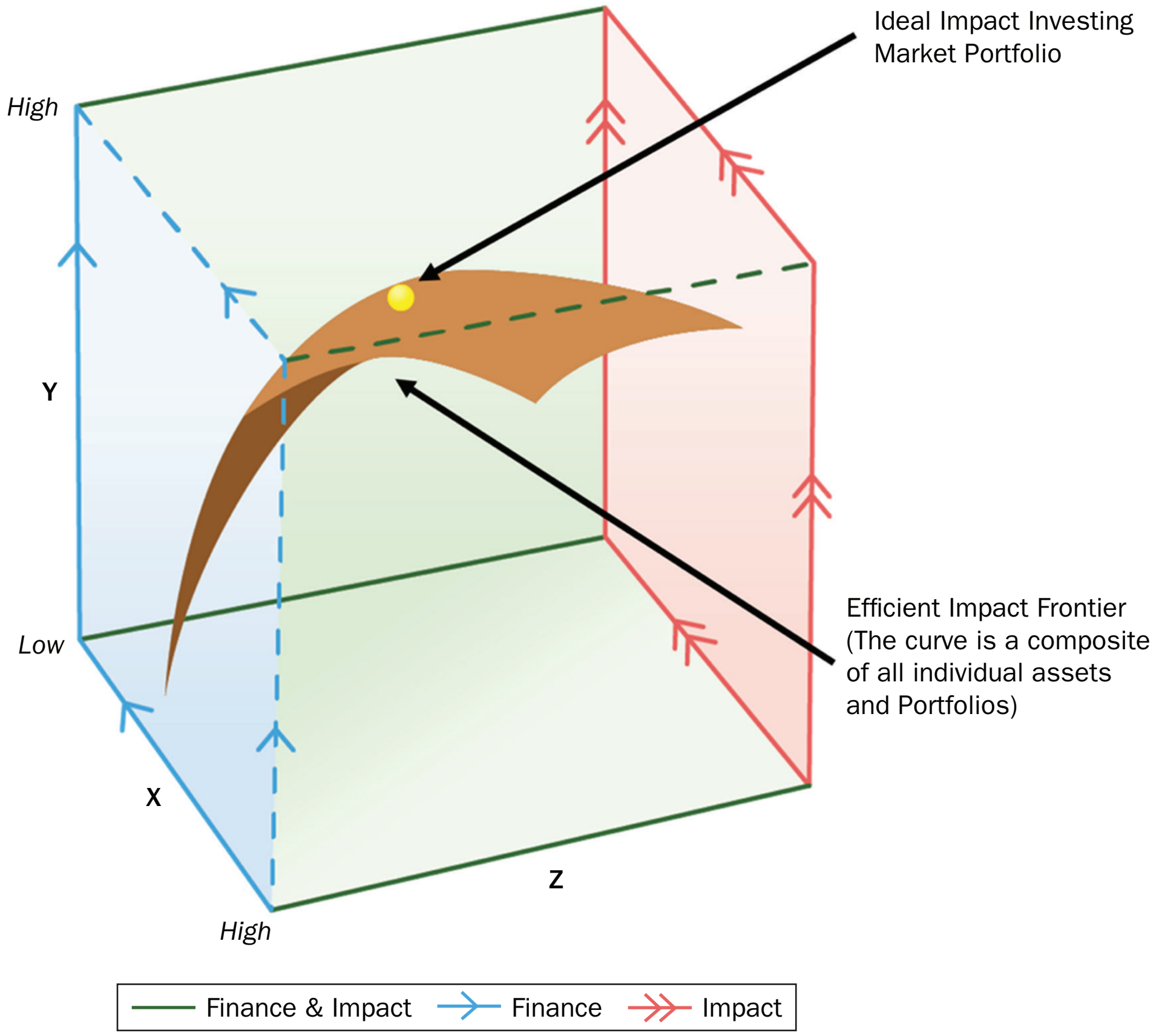

[14] It is important to remember that financial and impact risks and returns are distinct factors. Future research should focus on identifying potential correlations once a critical mass of impact data becomes available.

[15] Using this concept, the rate of return can be calculated for both financial and impact. This study will focus on calculating both financial and impact return combined and in a focused use case for monetized outcomes and impact.

As impact datasets standardize and scale, it will become more feasible to assess both financial and impact risks and returns, thereby allowing the establishment of minimum acceptable impact thresholds for every asset class and theme in impact investing. Impact IRR represents the first step in mapping EIR, focusing on monetized outcomes and impact (see "Initial Use Cases"). Future research may focus on use cases for non-monetized impact metrics in which monetization is unfeasible.

## Impact IRR/Impact NPV Rationale

IRR/NPV is widely regarded as one of the market standards for measuring an investment's potential return and evaluating an investment during the investment term. Net present value (NPV) is the present value of the cash flows at a project's required rate of return compared to one's initial investment (Gallo 2014). The internal rate of return (IRR) is the discount rate that makes a project's NPV zero. In other words, the expected compound annual rate of return will be earned on an investment (Vipond, n.d.).

IRR/NPV considers the following factors: time value of money, consideration of cash flows, comparison with cost of capital, accounting for reinvestment assumptions, ability to evaluate complex investments, uniform measure of comparison, and reflecting investor preference.

IRR/NPV produces a concrete number that even a novice investor can use to compare how an initial investment in one opportunity adds to the overall net monetary return relative to its alternatives (Gallo 2014).

How an investment's projected yield compares to that of another does not necessarily make it a better investment; the measure simply allows comparison with the required rate of return for the capital and is a tool to aid a final investment decision.

## Initial Use Cases

Impact IRR can be utilized for all impact investments in which an outcome or impact can be monetized.[16] The business sectors in which the initial impact IRR use cases shown in Exhibit 7 operate have datasets from which monetized outcomes and impact can be extracted or determined. The data must have the following characteristics (Barry 2024):[17]

### EXHIBIT 7
**Initial Impact IRR Use Cases**

| Sector | Housing | Small Business Investment | Healthcare |
|---|---|---|---|
| Outcome Objectives | Wealth creation/retention. Income stabilization. | Wealth creation. Jobs created/retained, job quality. | Improved health outcomes. |
| Datasets | ***Rental***: Rent rolls for affordable rents paid. Third-party data for comparable market-rate rents. ***For sale***: Mortgage amortization, housing valuesfrom third-party data providers. | **Real estate (purchase versus renting):** Estimated rents and real estate evaluation from third-party sources. **Wealth creation and job creation:** Primary datasets today are self-reported surveys. There is an opportunity to use nonidentifiable tax information for both business owner and business. | Nonidentifiable population or client-level data on decreased use of targeted medical facilities (e.g., ER visits) or health outcomes (reduced number of clients being transferred to nursing homes versus aging in community). |

[16] Impact theme sectors with monetized outcomes include energy, financial services, infrastructure, manufacturing, forestry and timber, water, and education.

[17] The data characteristics mentioned can also be used in conjunction with IF characteristics of useful information (Impact Frontiers 2024).

1. *Standardized*: To facilitate aggregation, all data collected from businesses have standardized measurement units and time frames.
2. *Verified*: Verify the accuracy and reliability of the data businesses provide, using third-party validation where possible.
3. *Confidential*: Maintain strict confidentiality of data and ensure that data sharing does not compromise a business's confidentiality.
4. *Current*: An agreed-upon cadence for updating data to reflect changes in business performance and corresponding client outputs, outcomes, and/or impact.

## Explanation of Impact IRR

Impact IRR/impact NPV has three categories of unique components. Some are unique to impact IRR/NPV, and some are shared with IRR/NPV:

### Components that Account for Both Impact and Financial Returns

1. *Term*: the time period of an investment, the provisions of an agreement or contract, and the lifespan assigned to an asset or liability (Kenton 2023).
2. *Hurdle rate*: the minimum rate of return on a project or investment required by a manager or investor (Kenton 2024). Embedded in the hurdle rate is the opportunity cost: the value an investor is willing to forgo as the result of making an investment (Fernando 2024).
3. *Catalytic opportunity*: classifying investments that can induce additional necessary capital to invest in impact-driven enterprises.

### Components that Account for Impact Returns

1. *Attribution and deadweight*: the percentage of social and/or environmental outcome(s) and impact that can be assigned to an investment.
2. *Impact variability*: the rate at which the desired social and/or environmental outcome(s) and impact may change throughout the lifetime of the project and/or investment.

### Components that Account for Financial Returns

1. *Total investment*: total projected investment into a given project/program.
2. *Projected financial return*: the financial gain or loss of an investment in a particular period.

### Unique IIRR/INPV Elements Overview

The IIRR/NPV formula has seven elements (see Exhibit 8). This section will use an example investment to explain each element and demonstrate how they will be calculated for the example investment (Exhibit 9), then solve for the impact IRR (making INPV = 0 and solving for $r$).

### EXHIBIT 8
**Impact Net Present Value Formula**

$$\text{INPV} = \sum_{t=1}^{T} \frac{C_t + I_t(C_o/D)}{(1+r)^t} - C_o$$

**Where:**

$T$ = Total number of periods
$t$ = Measurement of time period(s) (e.g., year, month)
$I$ = Projected impact (social and/or environmental) return for investment
$C_t$ = Projected financial return for investment
$r$ = Cost of capital/hurdle rate
$C_o$ = Total initial investment
$D$ = Total investment with similar terms in the capital stack (equity, debt, etc.)

### EXHIBIT 9
**Example Investment**

Fictitium Foundation (FF) is interested in making a $1.6 million PRI with an annualized return of 2% and a term of 10 years that will be invested to preserve a 42-unit naturally occurring affordable housing development. FF's $1.6 million PRI is the only BIC; all other financing is market-rate impact debt. Total investment = $4.2 million. FF's cost of capital is 6%.

**Investment Details**

Amount: $1.6 million PRI, subordinate debt annualized return, 2% (interest-only; capital returned as a balloon payment at the end of the term), 10-year term.

## Element 1: Term (T)

Using the IIRR/INPV formula as the basis of the calculation, an investor's impact investment is capped by the investment term (thereby capping the investor's impact benefit and financial benefit derived from an investment).[18]

**Example investment and graphic representation.** FF's investment has a term of 10 years. Therefore, $T = 10$ (see Exhibit 10).

### EXHIBIT 10
**Term Calculation**

$$\text{INPV} = \sum_{t=1}^{10} \frac{C_t + I_t(C_o/D)}{(1+r)^t} - C_o$$

| 0 | 1 | 2 | 3 | 4 | 5 | 6 | 7 | 8 | 9 | 10 |
|---|---|---|---|---|---|---|---|---|---|---|
| | | | | | | | | | | |

[18] Currently, there are early-stage explorations regarding whether concomitant outcomes during the investment period and long-term impacts post-investment that correlate with primary outcomes measured during the investment period can be attributed. Parallels exist in sectors such as public health and healthcare, in which studies have shown a strong correlation between interventions, concomitant outcomes, and long-term impact. Early testing focuses on impact investments in which products or services have robust research demonstrating a high correlation between concomitant outcomes and long-term impact to assess the extent to which attribution is reasonable and legally permissible. Once an approach is established, the impact IRR methodology can be updated.

### Element 2: Total Initial Investment ($C_o$)

When calculating the total investment, impact NPV classifies investments into two categories: (1) an investment in a fund or (2) a direct investment (see Exhibit 11):

1. *Investment in a fund*: An investor pools money with other investors into a fund that then invests in projects or enterprises.
   In typical cases, $C_o$ for an investor is obtained by calculating the proportion of a fund's total pooled money that has been committed to a specific investment and then applying that ratio to the total amount invested into the fund by that investor.
2. *Direct investment*: In this case, $C_o$ is an investor's total investment made directly into a project.

**Example investment and graphic representation.** FF has a total direct investment of **$1.6 million**. Therefore, **$C_o$ = $1.6 million** (see Exhibit 11).

## EXHIBIT 11

### Total Initial Investment Calculation

$$\text{INPV} = \sum_{t=1}^{10} \frac{C_t + I_t(1.6\text{M}/D)}{(1+r)^t} - 1.6\text{M}$$

| 0 | 1 | 2 | 3 | 4 | 5 | 6 | 7 | 8 | 9 | 10 |
|---|---|---|---|---|---|---|---|---|---|---|
| (1.6M) | | | | | | | | | | |

### Element 3: Projected Financial Return ($C_t$)

The projected financial return is the expected financial return received by the investor during the investment period plus the amount of initial investment capital that is expected to be returned at the end of the period. In the example, the project has a strong track record of returning capital to investors; thus, the investment capital returned is expected to equal $C_o$ (see Exhibit 12). The projected financial return must be adjusted accordingly if an investment has a higher risk profile.

**Example investment and graphic representation.** The FF investment is a **$1.6 million PRI investment** with **an annualized interest-only return of 2%** and **a term of 10 years** and **a balloon principal payment in year 10**. Therefore, the projected financial return, $C_t$, is **$1.6 million × .02 = $32,000/year + $1.6 million** (see Exhibit 12).

## EXHIBIT 12

### Projected Financial Return Calculation

$$\text{INPV} = \sum_{t=1}^{10} \frac{\overset{+\,1.6\text{M}}{32\text{k}} + S_t(1.6\text{M}/D)}{(1+r)^t} - 1.6\text{M}$$

| 0 | 1 | 2 | 3 | 4 | 5 | 6 | 7 | 8 | 9 | 10 |
|---|---|---|---|---|---|---|---|---|---|---|
| (1.6M) | 32k | 32k | 32k | 32k | 32k | 32k | 32k | 32k | 32k | 32k + 1.6M |

### Element 4: Projected Impact Return ($I_t$)

The social and environmental returns of an investment—the outputs, outcomes, and/or impacts (referred to collectively as "resultants")—are translated into a dollar amount that can be directly attributed to that investment. This concept is an emerging space in impact investing. It is essential to define the primary, secondary, and tertiary resultants that may occur. This level of analysis will inform impact investment decisions, define the markets, and indicate to prospective investors where BIC will be most beneficial.

The working definitions for outputs, outcomes, and impact are as follows:

- *Outputs*: the direct result of an invested business's activities, including their products, services, and any byproducts that can be measured by output indicators[19] or deliverables.[20]
- *Outcomes*: the projected short- and medium-term effects, both positive and negative, of the products or services provided by the investee business to the target clients or the benefits that thereby flow to the natural environment.[21]
- *Impact*: positive and negative primary and secondary long-term effects produced by the investee business's products or services that benefit the target clients or natural environment and can contribute to population-level change.[22]

One must also account for impact variability, defined as the rate at which a resultant may change through the lifetime of a project and/or investment (Social Value International, n.d.). Investors must assess this risk for each investment and adjust the impact residuals accordingly.

To assist in projecting the likelihood of the desired resultant, one can determine the level of evidence correlating with the likelihood that the resultant will be realized.[23]

1. *Scientific consensus*: Systematic reviews of the empirical evidence document a scientific consensus on the likelihood that resultant(s) will be realized.
2. *Empirical evidence*: Empirical studies show that the resultant(s) have been realized in specific settings. (Findings cannot yet, however, be generalized.)
3. *Model-based predictions*: Models predict that the resultant(s) should be realized under certain assumptions.
4. *Narrative*: An explanation that rationalizes why the resultant(s) may be effective.

The investor in the example does not project any variability. The investment is an established product type with a track record of benefiting target clients, and scientific

[19] An output indicator measures the immediate products or services produced by a program or project, offering information on the quantity and quality of goods or services created.

[20] Definition derived from Impact Frontier's output definition (Impact Management Platform 2021).

[21] Definition derived from Impact Frontier's outcome definition (Impact Management Platform 2021).

[22] Definition derived from the Organisation for Economic Cooperation and Development definition (Organisation for Economic Cooperation and Development 2021) and also referencing Impact Frontier's definition of impact (Impact Management Platform 2021); impact is unlikely within the time frames of most impact investments.

[23] Based on the definition of level of evidence given by Heeb and Kölbel (n.d.); other resources include the GIIN IRIS+ System and research from Value-Based Alliance and the International Foundation for Valuing Impacts.

consensus supports its projections that the impact created will extend beyond the investment term.

The project impact return created by an investment is the sum of the projected monetized impact returns for each benefit for each year of the investment term. It is essential to monitor the investment and adjust the projections as resultant data becomes available.

**Example investment and graphic representation.** FF's investment is projected to sustain 42 affordable housing units.

The net projected monthly impact benefit is $13,110. Given that the potential investment is a stable asset with a long track record, there is little expectation of impact variability. Therefore, the net projected annual impact return, $I_t$, is 12 × $13,110 = $157,315 (see Exhibit 13).[24]

## EXHIBIT 13

### Projected Impact Return Calculation

**Panel A:**

| | Rents | | | Rental Property Unit Distribution[b] | | | |
|---|---|---|---|---|---|---|---|
| **Income Levels** | **1B** | **2B** | **3B** | **1B** | **2B** | **3B** | **Total** |
| 30%[a] | $532 | $639 | $738 | 4 | 4 | 1 | 9 |
| 50%[a] | $888 | $1,066 | $1,231 | 10 | 12 | 3 | 25 |
| 60%[a] | $1,065 | $1,279 | $1,477 | 3 | 5 | 0 | 8 |
| Market Rate | $1,159 | $1,285 | $1,801 | 0 | 0 | 0 | 0 |
| Total | | | | ***17*** | ***21*** | ***4*** | ***42*** |

**Panel B: Gross Projected Monthly Impact**

| **Income Levels** | **1B** | **2B** | **3B** | **Total** |
|---|---|---|---|---|
| 30%[a] | $2,508 | $2,584 | $1,063 | $6,155 |
| 50%[a] | $2,710 | $2,628 | $1,710 | $7,048 |
| 60%[a] | $282 | $30 | – | $312 |
| Market Rate | – | – | – | – |
| Total | **$5,500** | **$5,242** | **$2,773** | **$13,515** |

**Panel C: Annual Projected Impact Return[a]**

| | | |
|---|---|---|
| **Gross Projected Monthly Impact** | $13,515 | |
| **Vacancy (based on 3% vacancy rate)** | ($405) | *–3%* |
| **Net Monthly Impact** | $13,110 | |
| **Net Annual Impact Return** | **$157,315** | *x12* |

NOTES: **Panel A:** [a]The rents are capped at 30% of gross income, following US Housing and Urban Development guidelines based on unit bedrooms and a household's maximum income (Office of Policy Development and Research, n.d.). [b]The unit distribution is based on the existing tenant income profiles. **Panel B:** [a]The projected monthly impacts are calculated using the rental property unit distribution. The difference between the market rate rent and affordable rent is multiplied by the number of units in the property. The projected impact calculations for each unit type based on affordability and bedroom size are then summed to calculate the gross projected monthly benefit. **Panel C:** [a]The net annual projected impact return is calculated by subtracting the projected vacancy from the gross projected monthly impact and multiplying the resulting net monthly impact benefit by 12 (see Exhibit 14).

[24] In the US affordable housing market, a key outcome is to promote income stability. A key component of income stability is to ensure households spend no more than 30% of their gross income on housing, measured by the difference between market-rate rent and affordable rent. If they spend more than 30%, they are deemed to be moderate or severely housing cost–burdened (Airgood-Obrycki, Hermann, and Wedeen 2024; Office of Policy Development and Research, n.d.).

## EXHIBIT 14

### Projected Impact Return Calculation

$$INPV = \sum_{t=1}^{10} \frac{\overset{+\,1.6M}{32k + 157k(1.6M/D)}}{(1+r)^t} - 1.6M$$

| 0 | 1 | 2 | 3 | 4 | 5 | 6 | 7 | 8 | 9 | 10 |
|---|---|---|---|---|---|---|---|---|---|---|
| (1.6M) | 32k + 157k | 32k + 157k | 32k + 157k | 32k + 157k | 32k + 157k | 32k + 157k | 32k + 157k | 32k + 157k | 32k + 157k | 32k + 157k 1.6M |

### Element 5: Hurdle Rate (*r*)

The minimum required rate of return an investment must meet to be deemed a good investment is the hurdle rate (Corporate Finance Institute, n.d.) or cost of capital (Knight 2015). The hurdle rate, *r*, is equal to the projected return of the capital if it were to be invested in the market in a comparable investment. Investors determine the hurdle rate they will use. The hurdle rate for both BIC and MIC is based on the following:

- *BIC*: A BIC investor is willing to forgo a market-rate return to produce an outcome or impact. The difference between the market rate and BIC returns represents the investor's opportunity cost. Therefore, the hurdle rate for BIC is calculated as the opportunity cost of not investing that capital in the market.[25]
- *MIC*: MIC has already priced its risk and return according to the market; thus, the hurdle rate is the investment's projected return.

**Example investment and graphic representation.** FF estimates that the projected rate of return for the $1.6 million invested in the market **is 6%**. Therefore, $r = .06$ (see Exhibit 15).

## EXHIBIT 15

### Hurdle Rate Calculation

$$INPV = \sum_{t=1}^{10} \frac{\overset{+\,1.6M}{32k + 157k(1.6M/D)}}{(1+.06)^t} - 1.6M$$

| 0 | 1 | 2 | 3 | 4 | 5 | 6 | 7 | 8 | 9 | 10 |
|---|---|---|---|---|---|---|---|---|---|---|
| (1.6M) | $\frac{32k + 157k}{1.06^1}$ | $\frac{32k + 157k}{1.06^2}$ | $\frac{32k + 157k}{1.06^3}$ | $\frac{32k + 157k}{1.06^4}$ | $\frac{32k + 157k}{1.06^5}$ | $\frac{32k + 157k}{1.06^6}$ | $\frac{32k + 157k}{1.06^7}$ | $\frac{32k + 157k}{1.06^8}$ | $\frac{32k + 157k}{1.06^9}$ | $\frac{32k + 157k + 1.6M}{1.06^{10}}$ |

[25] This represents an initial method of calculating the hurdle rate. It is meant as a starting point. Investors can utilize this method or develop their own rationale for valuing the opportunity cost of funds.

### Element 6: Attribution and Deadweight

Attribution is defined as the prorated share of impact return associated with the amount of the investor's capital within the same tier (see Exhibit 16).

Attribution is achieved by summing the discounted impact returns of an investment and then multiplying that sum by the value of that investment divided by the total amount invested in the same tier by the investor.

Deadweight is defined as the cost to society created by market inefficiency, which occurs when supply and demand are out of equilibrium (Tuovila 2024). Deadweight in the context of impact IRR/impact NPV is surplus investment capital in a particular investment tier; it is considered thus because it provides no further economic or impact returns. There may be noneconomic or impact reasons for additional investors (e.g., introducing a risk-averse investor to impact investing, or it is politically prudent). If an investment is deemed deadweight, the share of potential attributed impact return can be reduced, or the potential attributed impact return distributed pro rata among the investors in that tier.

**Example investment and graphic representation (attribution).** FF is interested in making a **$1.6 million** PRI that will be **the only BIC in the deal**. The investment is classified as Tier 1. ($C_o/D$) = Total FF investment/Total Tier 1 capital = **1.6 million/1.6 million** (see Exhibit 17).

**EXHIBIT 16**

**Tier Categories**

| Tier 1 | Tier 2 | Tier 3 |
|---|---|---|
| BIC debt and equity | Equity and equity-like investments, preferred stock | Debt and debt-like investments, common stock |

**EXHIBIT 17**

**Attribution and Deadweight Calculation**

$$\text{INPV} = \sum_{t=1}^{10} \frac{32k + 157k(1.6M/1.6M) \; {+\,1.6M}}{(1 + 0.06)^t} - 1.6M$$

| 0 | 1 | 2 | 3 | 4 | 5 | 6 | 7 | 8 | 9 | 10 |
|---|---|---|---|---|---|---|---|---|---|---|
| (1.6M) | $\frac{32k + 157k \times (1.6M/1.6M)}{1.06^1}$ | $\frac{32k + 157k \times (1.6M/1.6M)}{1.06^2}$ | $\frac{32k + 157k \times (1.6M/1.6M)}{1.06^3}$ | $\frac{32k + 157k \times (1.6M/1.6M)}{1.06^4}$ | $\frac{32k + 157k \times (1.6M/1.6M)}{1.06^5}$ | $\frac{32k + 157k \times (1.6M/1.6M)}{1.06^6}$ | $\frac{32k + 157k \times (1.6M/1.6M)}{1.06^7}$ | $\frac{32k + 157k \times (1.6M/1.6M)}{1.06^8}$ | $\frac{32k + 157k \times (1.6M/1.6M)}{1.06^9}$ | $\frac{32k + 157k \times (1.6M/1.6M) + 1.6M}{1.06^{10}}$ |

### Element 7: Catalytic Opportunity

Catalytic opportunity describes investments designed to generate or enhance impact by attracting market-rate investors who would not otherwise invest. There are two types of catalytic capital:

1. *Catalytic BIC*: Catalytic BIC includes guarantees for market-rate capital, debt at below-market rates, and equity with asymmetrical returns. By definition, BIC *is* catalytic, and most catalytic capital is BIC.

2. *Catalytic MIC*: Catalytic MIC includes financial instruments priced at market rates, sometimes with an added pricing adjustment/premium to offset the risk of being the first mover. Catalytic MIC investments are primarily made by market-rate investors who believe that the market is undervaluing an investment and are willing to invest if compensated for the additional risk of being first movers.

In addition to attracting additional investors to an impact-driven enterprise or project, catalytic capital may provide a proof point to realign risk and return assumptions for non-impact investors, thus unlocking a new source of capital and thereby eliminating the need for catalytic capital for future projects. Systemic change of this type is unlikely to occur with a single investment. A critical mass of catalytic investments will likely be required to provide enough proof to induce market-rate investors to change their behavior.[26]

It is possible for multiple investors in an impact investment structure to be deemed catalytic if the favorable terms of their investments have separately induced additional investors to invest in the deal. An impact investment deal does not require catalytic capital.

**Example investment (catalytic capital).** FF's **$1.6 million PRI is the only BIC in the deal**. Therefore, FF's investment is BIC and has a **catalytic opportunity**.

### Example Investment: Final Calculations and Recommendations

Once all the information is entered into the formula, the impact NPV and impact IRR are calculated. I also provide an example due diligence summary of the investment recommendation (see Exhibit 18).

## EXHIBIT 18

### Example Investment: Final Calculations and Recommendations

$$\text{INPV} = \sum_{t=1}^{10} \frac{\overset{+\,1.6M}{32k + 157k(1.6M/1.6M)}}{(1 + 0.06)^{t}} - 1.6M = 696k$$

$$\text{INPV} = 0,\ \text{Then:} \sum_{t=1}^{10} \frac{\overset{+\,1.6M}{32k + 157k(1.6M/1.6M)}}{(1 + r)^{t}} - 1.6M = \text{IIRR} = r = {\sim}12\%$$

| Due Diligence | |
|---|---|
| **Investment** | $1.6 million PRI, subordinate debt with a projected return of 2%, 10-year term. Interest only. Rate: 6%. |
| **Total Number of Units** | 42 (100% affordable) |
| **Type of Investment** | Housing, preserve naturally occurring affordable housing |
| **Level of Evidence** | Scientific consensus |
| **Attribution and Catalytic Opportunity** | Tier 1. Potential catalytic opportunity. |
| **Notable Outcomes** | Residents are projected to realize an aggregate annual net benefit of $157,000, resulting in an average annual savings per household of $3,700, or $37,000 per household on housing costs over the 10-year investment term. |
| **Projected Impact IRR/NPV** | 12%, $696,000 |
| **Recommendation** | **Consider for possible investment** |

[26] A systemic change of this nature is only possible if there is an opportunity to realign market-rate investors' perceptions of the actual risk and return for a specific type of impact investment. If there is no opportunity to realign risk and return, these impact investments will be viable only with BIC (i.e., BIC will be catalytic for the investment but not the sector).

## ILLUSTRATIVE USE CASES

The illustrative use cases are based on investments from three leading impact investors. Certain proprietary information has not been shared, and projections are made based on the information available at the time of writing.

### Illustrative Use Case: LISC

Local Initiatives Support Corporation (LISC) is an impact investment fund that supports community development in housing, small business, financial health/jobs, education, safety, and health. LISC works in all 50 states and has over $32 billion in debt and equity investments.

**Investment overview.** The potential $2.55 million loan will refinance the senior debt and provide funds for capital improvement projects for an affordable senior rental community owned by a nonprofit (see Exhibit 19).

The property is an existing 104-unit, affordable senior housing property with a long-term subsidy contract (see Exhibit 20). Residents can access free services (health screenings, prevention and education, continuing education, art, and nutrition classes). Affordable units are in high demand, and there is a long waitlist.

**Mission alignment.** The investment:

- Preserves 100 high-demand housing units, providing affordable rental options for seniors and individuals with disabilities.
- Strengthens partnership with an existing affordable housing provider serving special needs populations.

**Projected impact.** The calculations are computed using the US federal subsidy (see Exhibit 20). The projected total gross monthly outcome is the total gross monthly subsidy (TGMS) calculated by multiplying the number of units by the applicable subsidies and summing the results (see Exhibit 20).

## EXHIBIT 19
### Investment Assumptions

| | |
|---|---|
| **Amount** | $2,545,000, 14 years, 4.25%, Amortizing |
| **Asset Class** | Real Estate |
| **Attribution** | Tier 1, Catalytic |
| **Impact Variability Rating** | 1, Scientific Consensus |
| **Annual Debt Service** | $244,926 |

## EXHIBIT 20
### Monthly Gross Monthly Subsidy

| | Unit Type | | | | | | | |
|---|---|---|---|---|---|---|---|---|
| | Studio | | One-Bedroom | | Two-Bedroom | | Total | |
| **Area Median Income** | **#** | **Subsidy** | **#** | **Subsidy** | **#** | **Subsidy** | **#** | **Subsidy** |
| *Extremely Low-Income* 30% | 2 | $686 | 47 | $825 | 1 | $1,200 | 50 | $2,711 |
| *Very Low-Income* 50% | 1 | $579 | 40 | $689 | 4 | $814 | 45 | $2,082 |
| *Low-Income* 80% | 1 | $427 | 4 | $396 | – | | 5 | $823 |
| **Total** | **4** | **$2,378** | **91** | **$67,911** | **5** | **$4,455** | **100** | **$74,744** |

TGMS is annualized, and vacancy is subtracted, giving a $834,143 projected net annual outcome (see Exhibit 21). The projected outcome is expected to increase by 3% annually.[27]

**Impact IRR.** The investment projects preserve 100 affordable units, yielding a $10.9 million outcome, 4.35% IRR, and 45% impact IRR (see Exhibits 22 and 23). LISC could include the free services as additional output metrics.

## EXHIBIT 21

### Projected Annual Outcome

| **TGMS** | **$74,744** |
|---|---|
| Total Gross Annual Subsidy | $896,928 |
| Vacancy* | ($62,785) |
| **Projected Net Annual Outcome** | **$834,143** |

**NOTE:** * 7% per underwriting criteria.

## EXHIBIT 22

### Projected Returns

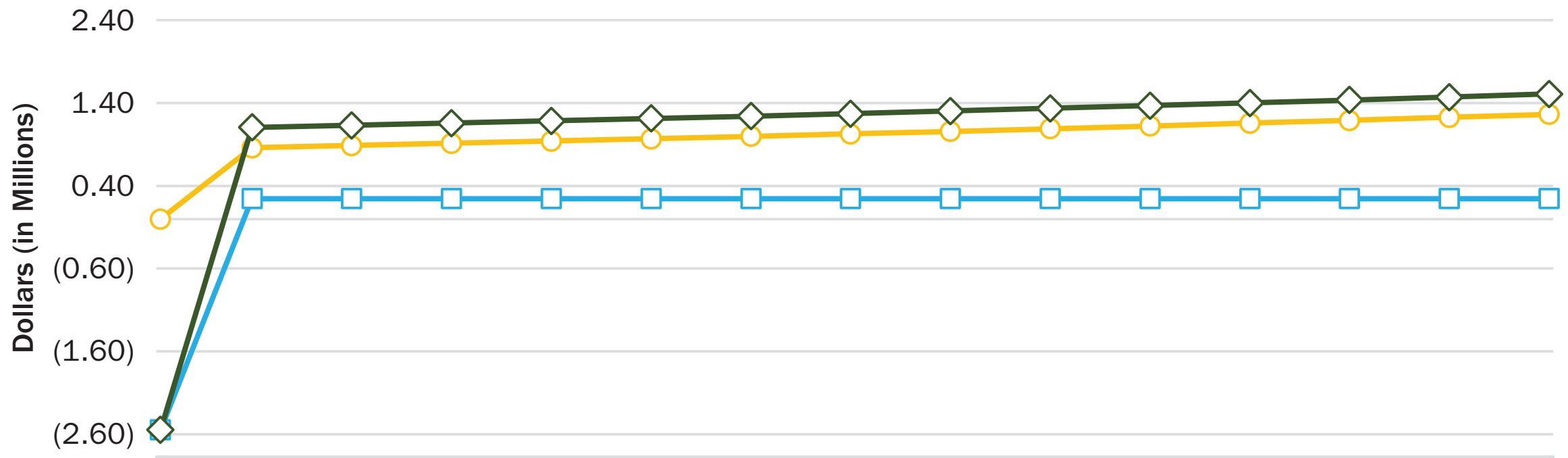

| | 01/22 | 12/22 | 12/23 | 12/24 | 12/25 | 12/26 | 12/27 | 12/28 | 12/29 | 12/30 | 12/31 | 12/32 | 12/33 | 12/34 | 12/35 |
|---|---|---|---|---|---|---|---|---|---|---|---|---|---|---|---|
| Financial Returns | (2.55) | 0.24 | 0.24 | 0.24 | 0.24 | 0.24 | 0.24 | 0.24 | 0.24 | 0.24 | 0.24 | 0.24 | 0.24 | 0.24 | 0.24 |
| Monetized Outcomes | – | 0.86 | 0.88 | 0.91 | 0.94 | 0.97 | 1.00 | 1.03 | 1.06 | 1.09 | 1.12 | 1.15 | 1.19 | 1.22 | 1.26 |
| Total Returns | (2.55) | 1.10 | 1.13 | 1.16 | 1.18 | 1.21 | 1.24 | 1.27 | 1.30 | 1.33 | 1.37 | 1.40 | 1.43 | 1.47 | 1.51 |

## EXHIBIT 23

### Due Diligence

$$\text{INPV} = 0 = \sum_{t=1}^{14} \frac{245\text{k} + 834\text{k}(2.545\text{M}/2.545\text{M})}{(1 + r)^t} - 2.545\text{M} = \text{IIRR} = r = \sim 45\%$$

| | |
|---|---|
| **Investment** | $2.55 million senior debt with projected return of 4.35%, amortizing |
| **Asset Class** | Real estate, affordable housing, 104 total: 100 affordable (96%), 4% market rate |
| **Additional Benefits to Target Clients** | Free education, financial health, and health services |
| **Attribution & Catalytic Opportunity** | Scientific consensus; Tier 1; catalytic opportunity |
| **Notable Outcomes** | Residents in affordable units are projected to realize an aggregate annual net benefit of $2 million, resulting in an average annual savings per household of $20,000, or $280,000 per household over the 14-year investment term |
| **Projected IRR/Projected Impact IRR/Total Outcome** | 4.25%, 45%, $10.9 million |
| **Recommendation** | **Consider for possible investment** |

[27] Three percent is commonly used when underwriting such investments.

It is important to note that the previous foundation example (FE) and the LISC case study are merely two data points. One should not infer that they are benchmarks[28]—they represent different types of investors with varying goals. Specifically, the differences in the impact IRR returns stem from three primary variables: investor hurdle rate (FE 6% versus LISC 4.25%), debt per unit (FE $38,000 versus LISC $25,000), and projected outcome per unit (FE $3,700 versus LISC $20,000).

### Illustrative Use Case: FFCP

Founders First Capital Partners (FFCP) provides capital to socially and economically disadvantaged small businesses, such as those led by people of color, women, veterans, or LGBTQ+ individuals or in low- or moderate-income areas. FFCP supports companies not typically served by venture capital or unable to access asset-based lending.

**Investment overview and impact alignment.** The US Small Business Administration reports that 4.7 million out of 8 million small business owners are underserved by traditional growth capital (Federal Reserve Banks 2022).

FFCP provides flexible and patient private debt to diverse-led small businesses aiming to grow into midmarket firms. This analysis focuses on FFCP's revenue-based financing (RBF) product, which is growth capital that offers flexible repayment such as equity but is nondilutive and does not require an exit like debt does (Catalyze 2024). RBF loans range from $50,000 to $2 million. FFCP has secured two tiers of debt for RBF (see Exhibit 24).

FFCP's impact measurement framework includes jobs created/sustained and job quality.

**Projected impact.** The impact metrics are monetized outcomes based on the dollar value of each job created and sustained by each loan. Years 1–3 are based on FFCP past performance; years 4–7 are projections (see Exhibits 25 and 26).

**Impact IRR.** The investments are projected to create the following:

- DT1: 348 jobs valued at $24 million with 7% IRR and 31% impact IRR
- DT2: 142 jobs valued at $12.5 million with 7% IRR and 53% impact IRR

The variation in impact IRR is due to DT1 investing in the fund's ramp-up phase (see Exhibits 27–29).

**EXHIBIT 24**

**Investment Assumptions**

| | Timing | Amount | Term | Attribution | Impact Variability Rating | Interest Rate |
|---|---|---|---|---|---|---|
| **Debt Tier 1 (DT1)** | Years 1–7 | $12 million | 7 years.<br>3 years interest-only,<br>4 years amortizing | Tier 1, Catalytic | 2, empirical evidence | 7% |
| **Debt Tier 2 (DT2)** | Years 4–7 | $8 million | 4 years amortizing | Tier 1, catalytic | 2, empirical evidence | 7% |
| **Total** | | **$20 million** | | | | |

[28] Future research will create benchmarks from historical performance to assess investments as impact data becomes available.

## EXHIBIT 25

### Loan and Impact Inputs

| | Archetype[a] | Business Contractors | Professional Business Services | Tach-Enabled Services | Tech/SaaS |
|---|---|---|---|---|---|
| **Years 1–3** | # of Loans[b] | 13 | 8 | 9 | 10 |
| | Average Loan | $350,000 | $200,000 | $325,000 | $310,000 |
| | New Jobs/$100k Invested[c] | $125,000 | $150,000 | $200,000 | $200,000 |
| **Years 4–7** | # of Loans Closed/Year[d] | 16 | 9 | 12 | 13 |
| | YOY Average Loan Amount Increase[e] | 7.50% | | | |

**NOTES:** [a]Business contractors: asset-light business services. Professional business services: asset-light professional services that support primary business activities. Tach-enabled services: tech-driven firms that streamline operations and support technological needs. Tech/SaaS: scalable technology companies that utilize subscription models. [b]Average term 12 months; based on FFCP past performance and initial data collected. [c]Based on preliminary results. The average total compensation for a new job created/sustained is $75,000, with costs increasing 3% annually. [d]Average term 12 months; years 4–7: 75% returning borrowers; 25% new borrowers. Returning borrowers' loan amounts will increase by 30% over four years; projected based on pent-up demand in the investment pipeline. [e]Years 4–7: 75% returning borrowers; 25% new borrowers. Returning borrowers' loan amounts will increase by 30% over four years; Projected based on pent-up demand in the investment pipeline.

## EXHIBIT 26

### Impact Projections

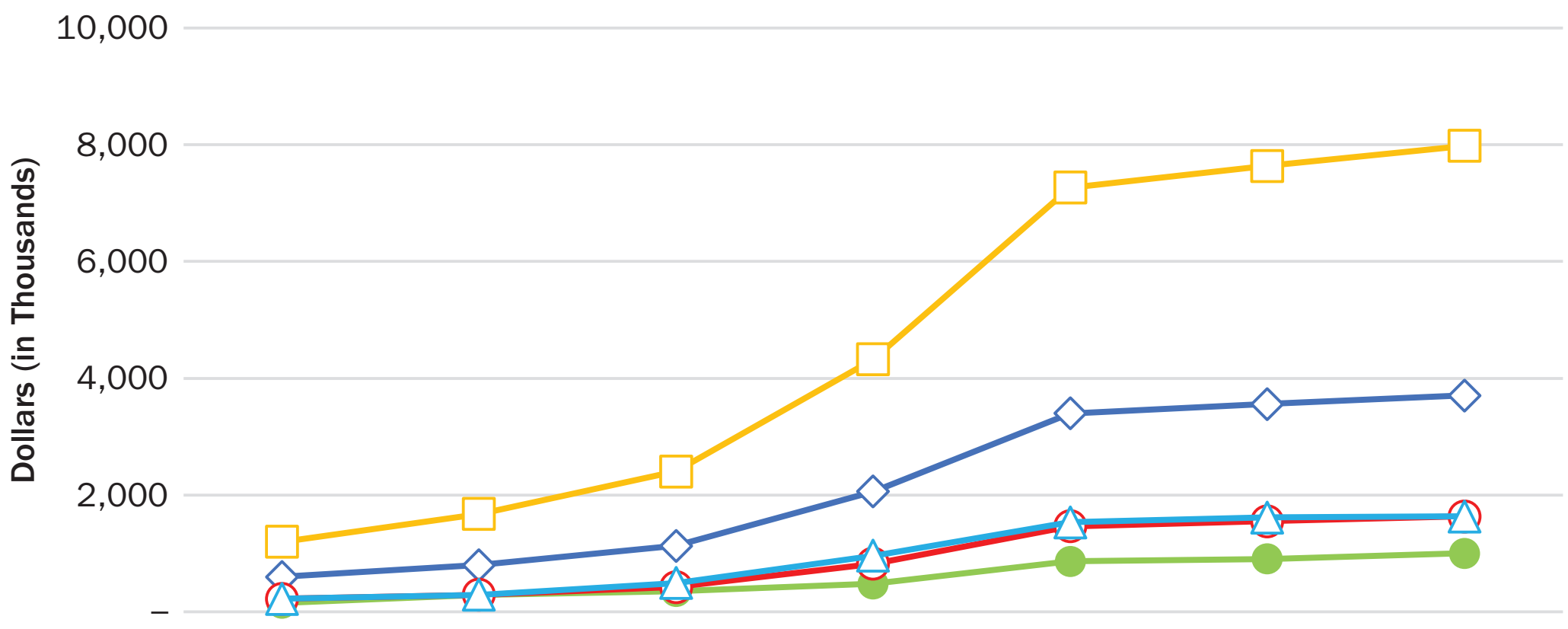

| | 12/19 | 12/20 | 12/21 | 12/22 | 12/23 | 12/24 | 12/25 |
|---|---|---|---|---|---|---|---|
| Business Contractors | 600 | 801 | 1,131 | 2,059 | 3,398 | 3,558 | 3,706 |
| Professional Business Services | 150 | 291 | 353 | 480 | 866 | 906 | 1,005 |
| Tech-Enabled Services | 225 | 291 | 424 | 824 | 1,466 | 1,553 | 1,633 |
| Tech/SAAS | 225 | 291 | 495 | 961 | 1,533 | 1,617 | 1,633 |
| Total | 1,200 | 1,675 | 2,404 | 4,324 | 7,263 | 7,634 | 7,977 |

## EXHIBIT 27

### Projected Returns

| | $, Term | Attribution | Impact Variability Rating | Jobs Created (#) | Jobs Created ($) | IRR | Impact IRR |
|---|---|---|---|---|---|---|---|
| **DT1** | $12 million, 7 years | Tier 1, catalytic | 2, empirical evidence | 348 | $24 million | 7% | 31% |
| **DT2** | $8 million, 3 years | Tier 1, catalytic | 2, empirical evidence | 142 | $12.5 million | 7% | 53% |

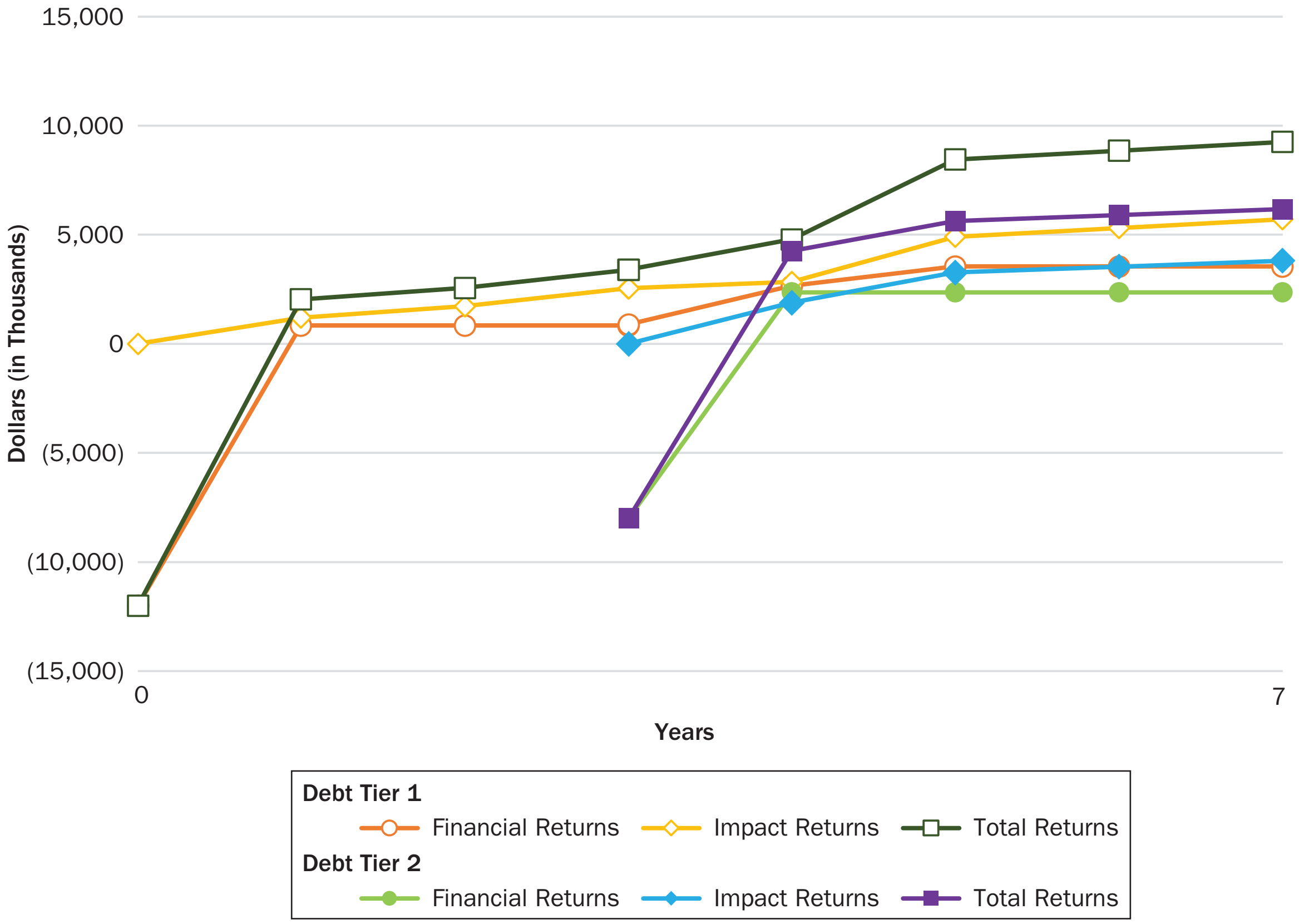

## EXHIBIT 28

### DT1 Projected Returns

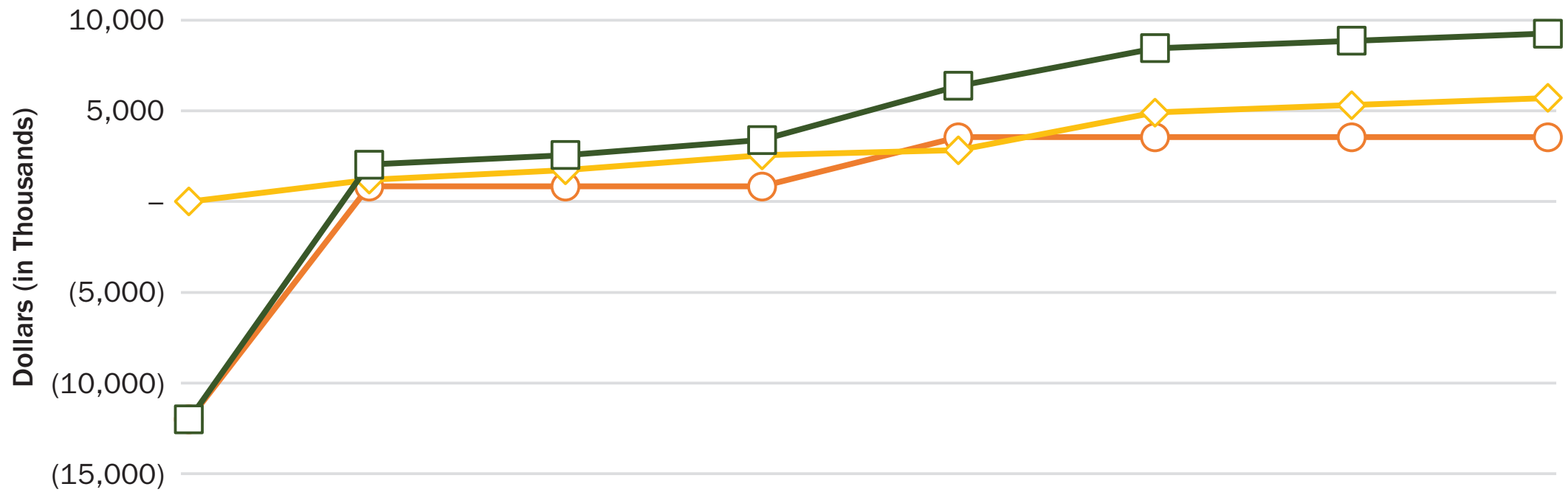

| | 01/19 | 12/19 | 12/20 | 12/21 | 12/22 | 12/23 | 12/24 | 12/25 |
|---|---|---|---|---|---|---|---|---|
| Financial Returns | (12,000) | 840 | 840 | 840 | 3,543 | 3,543 | 3,543 | 3,543 |
| Impact Returns | – | 1,200 | 1,725 | 2,550 | 2,835 | 4,905 | 5,310 | 5,715 |
| Total Returns | (12,000) | 2,040 | 2,565 | 3,390 | 6,378 | 8,448 | 8,853 | 9,258 |

$$\text{INPV} = 0 = \sum_{t=1}^{7} \frac{\text{FR} + \text{IR}^{1\text{–}3}(12\text{M}/12\text{M}) + \text{IR}^{4\text{–}7}(12\text{M}/20\text{M})}{(1+r)^t} - 12\text{M} = \text{IIRR} = r = {\sim}31\%$$

$T$ = 7 Years
$t$ = 1
$C_t$ = FR = Annual Financial Returns
$C_o$ = 12M

| | *Years 1–3* | *Years 4–7* |
|---|---|---|
| $I_t$ = | IR = Impact Returns | IR = Impact Returns |
| $C_o/D$ = | 12M/12M | 12M/20M |

## EXHIBIT 29

### DT2: Projected Returns

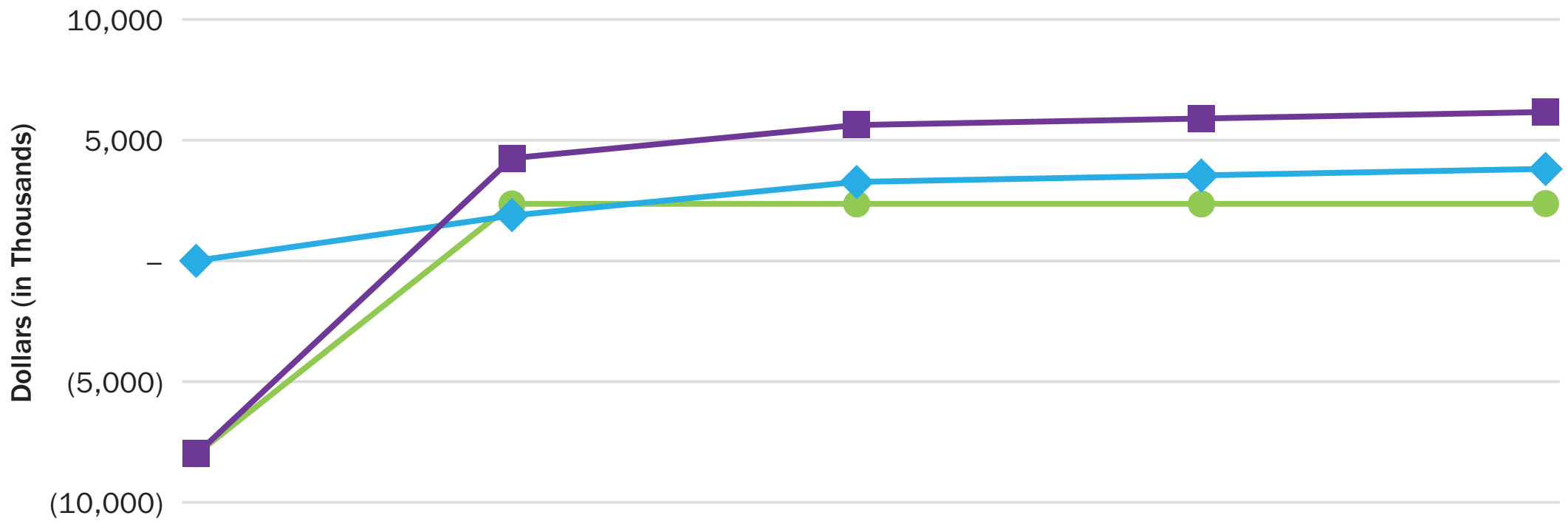

| | 01/22 | 12/22 | 12/23 | 12/24 | 12/25 |
|---|---|---|---|---|---|
| Financial Returns | (8,000) | 2,362 | 2,362 | 2,362 | 2,362 |
| Impact Returns | – | 1,890 | 3,270 | 3,540 | 3,810 |
| Total Returns | (8,000) | 4,252 | 5,632 | 5,902 | 6,172 |

$$\text{INPV} = 0 = \sum_{t=1}^{4} \frac{\text{FR} + \text{IR}\,(8\text{M}/20\text{M})}{(1+r)^t} - 8\text{M} = \text{IIRR} = r = 53\%$$

$T$ = 4 Years
$t$ = 1
$C_t$ = FR = Annual Financial Returns
$C_o$ = 8M
$S_t$ = IR = Impact Returns
$C_o/D$ = 8M/20M

### Illustrative Use Case: Learn Capital

Learn Capital (Learn) is a venture capital fund that invests in rapidly scaling tech-enabled education companies that empower individuals with the capacity, knowledge, and tools to thrive. Learn's focus areas include innovations in K–12 education, higher education, and lifelong learning. Learn has invested over $1.3 billion across 200-plus portfolio companies, reaching one billion learners. Learn is currently piloting a new impact methodology, the Learning Impact Index (LII), to measure the socio-economic value of its investments for individuals and systems.

**Investment overview and impact alignment.** The $2 million investment is for a startup online university (OU) that provides accredited online education for in-demand and future-focused skills, using proprietary technology to pair automation with human support (see Exhibit 30).

OU operates B2C and B2B2C models in 100-plus countries, primarily in low-income and middle-low-income countries (World Bank Group 2024). A student sample set of OU results demonstrates that a graduate's income increases significantly in the first two years (see Exhibit 31).

## EXHIBIT 30

### Investment Assumptions

| | |
|---|---|
| **Investment** | $2 Million, Series A |
| **Hurdle Rate** | 20% |
| **Total Series A** | $8 Million |
| **Attribution** | Tier 1, Catalytic |
| **Impact Variability Rating** | 2, Empirical Evidence |
| **Projected Gain from Sale in Year 10*** | $18 Million |

**NOTE:** * Based on a projected $300 million valuation in 2029, assuming 51% year-over-year growth.

## EXHIBIT 31

### Impact Assumptions[a]

| | |
|---|---|
| Average course length (months) | 24 |
| Graduate and nongraduate starting salary | $1,900 |
| Nongraduate average annual salary increase | 5% |
| **Graduate Income Assumptions** | |
| Year 1 salary increase | 30% |
| Year 2 salary increase (versus base year) | 55% |
| Year 3+ salary increase | 5% |
| **Program Cost Assumptions** | |
| Cost | $2,339 |
| Graduates: self-financed (paid in year course completed) | $234 |
| Graduates: total financed | $2,105 |
| Graduates: annual debt service (5%, 6-year amortizing loan) | $356 |
| Resignations (i.e., drop-outs) | $234 (10%) |
| Annual cost increase | 3% |
| Average scholarships and employer-paid (%) | 25% |

**NOTES:** [a]The graduates retain all additional income as a result of completing the course, reported either through verified self-reports or by the employer who funded the course. The business model does not employ any income-sharing agreements with graduates, whether with the investor, investee, or employer.

Learn is piloting LII with OU since there is a measurable direct economic benefit to clients.

**Projected impact.** The model assumes increases in graduate net income over the investment period based on OU's current outcomes and growth projections (see Exhibits 31 and 32).

The annual increase in net income (see Exhibit 33) is calculated thus:

1. *Change in net income of graduates*: Subtract the self-financed amount and the comparable nongraduate income from the graduate income, then multiply that figure by the number of graduates.
2. *Resigners' repayments*: Subtract from the total number of program resigners multiplied by their repayment.
3. *Impact attribution*: Multiply the resulting figure by 0.25 ($2 million Learn's investment/$8 million Series A total).

**Impact IRR.** The impact projects are calculated to be 25% IRR and 29% impact IRR (see Exhibits 34 and 35). As OU expands its presence in middle-income countries, the project outcomes and impact return will likely increase.

## EXHIBIT 32
### Growth Projections

| Year | 1 | 2 | 3 | 4 | 5 | 6 | 7 | 8 | 9 | 10 |
|---|---|---|---|---|---|---|---|---|---|---|
| Students | 160 | 513 | 850 | 1,363 | 2,058 | 3,107 | 4,692 | 7,085 | 10,699 | 16,155 |
| Graduates | 112 | 359 | 595 | 954 | 1,441 | 2,175 | 3,285 | 4,960 | 7,489 | 11,309 |

**NOTE:** * Actual figures years 1–5. ** Projections based on 51% year-over-year increase in student enrollment.

## EXHIBIT 33
### Projected Impact Returns

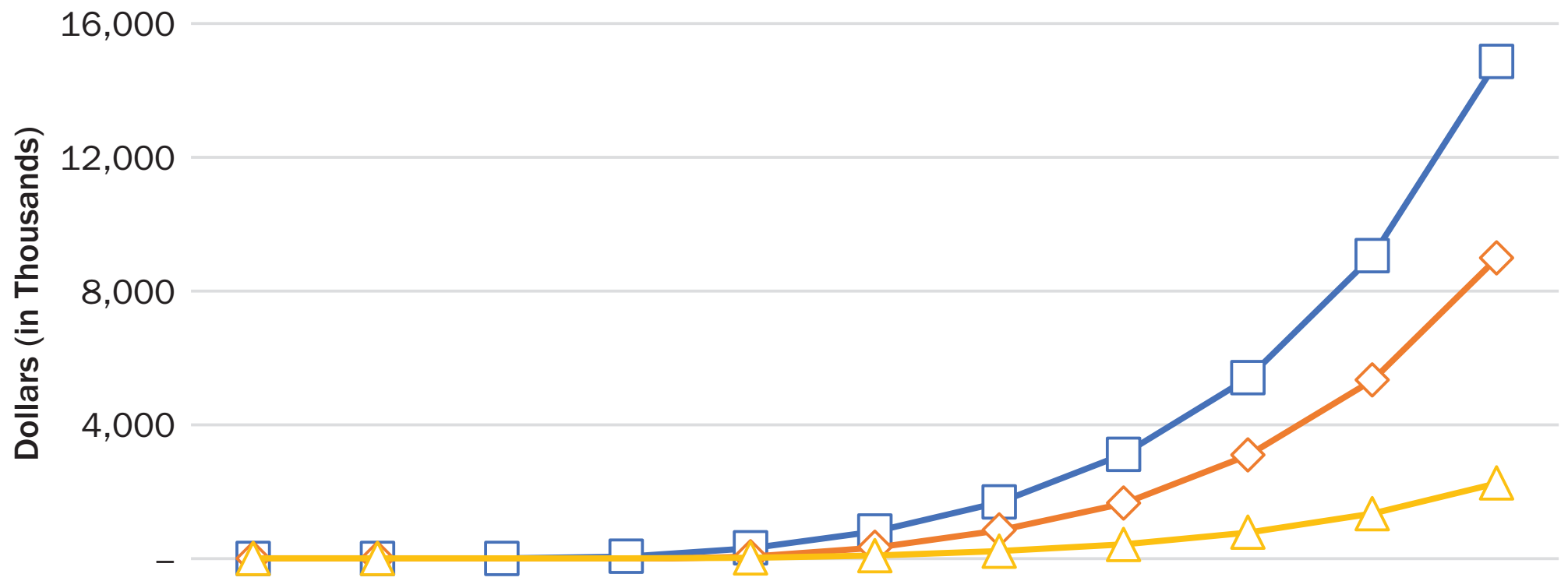

| | 12/19 | 12/20 | 12/21 | 12/22 | 12/23 | 12/24 | 12/25 | 12/26 | 12/27 | 12/28 | 12/29 |
|---|---|---|---|---|---|---|---|---|---|---|---|
| Graduate Gross Income | – | – | – | 59 | 302 | 807 | 1,680 | 3,109 | 5,397 | 9,048 | 14,861 |
| Graduate Net Income | – | – | (11) | (28) | 46 | 324 | 839 | 1,644 | 3,090 | 5,329 | 8,980 |
| Graduate Net Income Attributed to Learn | – | – | (3) | (7) | 11 | 81 | 210 | 411 | 773 | 1,332 | 2,245 |

## EXHIBIT 34

### Projected Investment Results

| Details | Attribution | Impact Variability Rating | Financial and Impact Projections | | | | |
|---|---|---|---|---|---|---|---|
| $12M, Series A | Tier 1, Catalytic | 2, Empirical Evidence | Financial: year 10 sale | Outcome: gross, net | Outcome attributed to Learn | IRR | Impact IRR |
| | | | $300 million; $18 million to Learn | $35.2 million, $20.2 million | $2.45 million | 25% | 29% |

## EXHIBIT 35

### Projected Returns

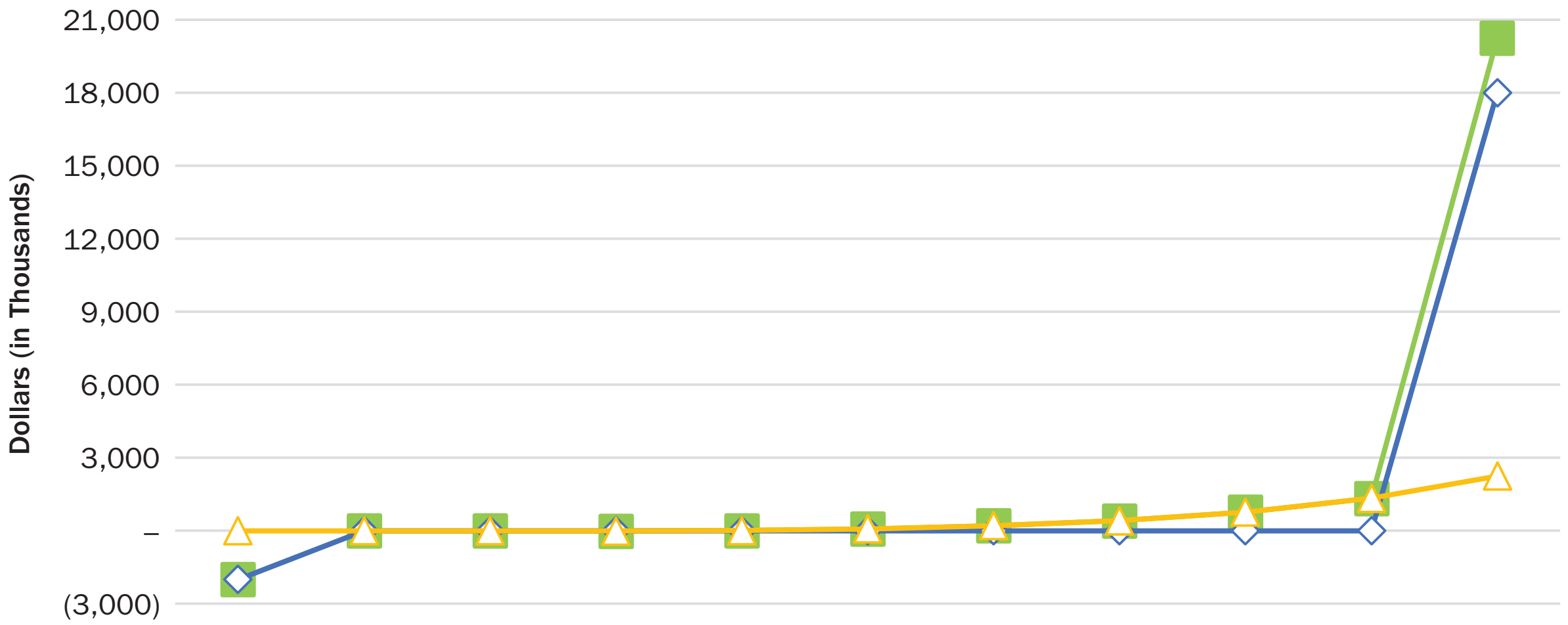

| | 12/19 | 12/20 | 12/21 | 12/22 | 12/23 | 12/24 | 12/25 | 12/26 | 12/27 | 12/28 | 12/29 |
|---|---|---|---|---|---|---|---|---|---|---|---|
| Total | (2,000) | – | (3) | (7) | 11 | 81 | 210 | 411 | 773 | 1,332 | 20,245 |
| Financial Returns | (2,000) | – | – | – | – | – | – | – | – | – | 18,000 |
| Impact Returns | – | – | (3) | (7) | 11 | 81 | 210 | 411 | 773 | 1,332 | 2,245 |

$$\text{INPV} = 0 = \sum_{t=1}^{10} \frac{18\text{M} + \text{IR}(2\text{M}/8\text{M})}{(1 + r)^t} - 2\text{M} = \text{IIRR} = r = 29\%$$

$T$ = 10 Years

$t$ = 1

$C_t$ = 18M (Year 10)

$S_t$ = "IR" = Impact Returns From Projections

$C_o/D$ = 2M/8M

$C_o$ = 2M

## CONCLUSION AND FUTURE WORK

This study represents the first steps in developing a scalable methodology for defining the impact markets. I developed and demonstrated a tool for monitoring and evaluating impact investments, utilizing established financial theories, leading impact thought leadership, and existing datasets.

Considering the various ways impact portfolios are currently constructed, it will be critical to integrate this methodology into existing and future portfolios to incorporate additional quantitative and qualitative impact metrics, regulatory filings, and disclosures to ensure all stakeholders can determine impact portfolio risk–return.

Work will continue through rigorous testing through an iterative process and will focus on the following:

1. Defining baseline outputs, outcomes, and impact for all use cases. Also, establishing how to collect, monitor, and evaluate data effectively and efficiently.
2. Refining attribution and deadweight metrics once there is a representative sample dataset.
3. Valuing projected outcomes and impact appropriately based on the level of evidence, as well as correlating concomitant outcomes and long-term impact.
4. Incorporating nonmonetizable outputs, outcomes, and impact valued by stakeholders alongside impact IRR.
5. Establishing an impact data repository or building on existing datasets to determine benchmarks and define impact markets.

## REFERENCES

Acumen. n.d. "Patient Capital." Acumen.

Airgood-Obrycki, W., A. Hermann, and S. Wedeen. 2024. *Deteriorating Rental Affordability An Update on America's Rental Housing 2024*. Cambridge: Joint Center for Housing Studies of Harvard University.

Barber, B., A. Morse, and A. Yasuda. 2021. "Impact Investing." *Journal of Financial Economics* 139 (1): 162–185.

Barry, M. 2024. "Partner, Learn Capital." Interview by D. Soliman.

Berk, J. B., and J. H. van Binsbergen. 2021. "The Impact of Impact Investing." Working paper, Stanford University Graduate School of Business.

BlueMark, and Morgan, Lewis & Bockius LLP. 2021. "Making Sense of Sustainable Investing: What Asset Managers Should Know About Compliance with Financial Regulations and Alignment with Industry Standards." BlueMark.

Broccardo, E., O. Hart, and L. Zingales. 2021. "Exit vs. Voice." Working paper, European Corporate Governance Institute.

Catalyze. 2024. "The State of Revenue—Based Financing & CDFIs." Catalyze.

Chowdhry, B., S. Davies, and B. Waters. 2018. "Investing for Impact." *Review of Financial Studies* 32 (3): 864–904.

Cole, S., M. Melecky, F. Mölders, and T. Reed. 2020. "Long-Run Returns to Private Equity in Emerging Markets." Working paper, World Bank Group.

Convergence. 2018. "Leverage of Concessional Capital." Convergence.

Corporate Finance Institute. n.d. "Hurdle Rate Definition." Corporate Finance Institute.

Edmans, A., D. Levit, and J. Schneemeier. 2023. "Socially Responsible Disinvestment." Working paper, European Corporate Governance Institute.

Federal Reserve Banks. 2022. "2022 Report on Employer Firms: Based on the 2021 Small Business Credit Survey." FED Small Business.

Feor, L., A. Clarke, and I. Dougherty. 2023. "Social Impact Measurement: A Systematic Literature Review and Future Research Directions." *World* 4 (4): 816–837.

Fernando, J. 2024. "Opportunity Cost: Definition, Formula, and Examples." Investopedia.

Gallo, A. 2014. "A Refresher on Net Present Value." Harvard Business Review.

Global Impact Investing Network. 2019. *Core Characteristics of Impact Investing*. New York: Global Impact Investing Network.

——. n.d. "Blended Finance Working Group." Global Impact Investing Network.

Goetzmann, W. N. n.d. *An Introduction to Investment Theory*. New Haven: Yale University.

Green, D., and B. Roth. 2024. "The Allocation of Socially Responsible Capital." *The Journal of Finance* 80 (2): 755–781.

Gupta, D., A. Kopytov, and J. Starmans. 2024. "The Pace of Change: Socially Responsible Investing in Private Markets." SSRN.

Hand, D., S. Sunderji, M. Ulanow, R. Remsberg, and K. Xiao. 2024. *State of the Market 2024: Trends, Performance and Allocations*. New York: Global Impact Investing Network.

Hand, D., M. Ulanow, H. Pan, and K. Xiao. 2024. *Sizing the Impact Investing Market 2024*. New York: Global Impact Investing Network.

Heeb, F., and J. Kölbel. n.d. *The Investor's Guide to Impact*. Zurich: University of Zurich.

Impact Frontiers. 2024. "Impact Performance Reporting Norms for Investors in Private Markets Version 1." Impact Frontiers.

——. n.d. "The Efficient Impact Frontier." Impact Frontiers.

Impact Management Platform. 2021. "Key Terms and Concepts." Impact Management Platform.

——. n.d. "Impact and the Impact Pathway." Impact Management Platform.

Internal Revenue Service. 2023. "Program Related Investments." Internal Revenue Service.

J. P. Morgan Asset Management. 2024. "The UK Sustainability Disclosure Requirements (UK SDR) Explained." J. P. Morgan Asset Management.

Jeffers, J., E. Lyu, and K. Posenau. 2024. "The Risk and Return of Impact Investing Funds." *Journal of Financial Economics* 161: 103928.

Kenton, W. 2023. "Term." Investopedia.

——. 2024. "Hurdle Rate: What It Is and How Businesses and Investors Use It." Investopedia.

Knight, J. 2015. *HBR Tools Return on Investment (ROI)*. Cambridge: Harvard Business Review.

Landier, A., and S. Lovo. 2020. *ESG Investing: How to Optimize Impact?* Paris: HEC Paris.

Lo, A., and R. Zhang. 2023. "Quantifying the Impact of Impact Investing." *Management Science* 70 (10): 6483–7343.

MacArthur Foundation. n.d. "Catalytic Capital Consortium." MacArthur Foundation.

Organisation for Economic Cooperation and Development. *Applying Evaluation Criteria Thoughtfully*. Paris: OECD.

Oehmke, M., and M. Opp. 2024. "A Theory of Socially Responsible Investment." *Review of Economic Studies* 92 (2): 1193–1225.

Office of Policy Development and Research. n.d. "HUD User." Income Limits.

Schlütter, D., L. Schätzlein, R. Hahn, and C. Waldner. 2024. "Missing the Impact in Impact Investing Research—A Systematic Review and Critical Reflection of the Literature." *Journal of Management Studies* 61 (6): 2694–2718.

Social Value International. n.d. "The Purpose of the Principles of Social Value and the SVI Standards." Social Value UK.

The Brainy Insights. 2024. *Impact Investing Market Size by Sector (Education, Agriculture, Healthcare, Energy, Housing, and Others), Investor (Individual Investors, Institutional Investors, and Others), Regions, Global Industry Analysis, Share, Growth, Trends, and Forecast 2024 to 2033*. New York: The Brainy Insights.

Tuovila, A. 2024. "What Is Deadweight Loss, How It's Created, and Economic Impact." Investopedia.

Vipond, T. n.d. "Internal Rate of Return (IRR)." Corporate Finance Institute.

Wagstaff, E., R. Revesz, and W. Stewart. 2024. "Screenings and Exclusions." Principles for Responsible Investment.

World Bank Group. 2024. "World Bank Income Groups." Our World in Data.

Zhou, M. 2022. "ESG, SRI, and Impact Investing: What's the Difference?" Investopedia.